\documentclass[10pt, journal, letterpaper]{IEEEtran}

\usepackage[T1]{fontenc}
\usepackage[cmex10]{amsmath}
\usepackage{amssymb}
\usepackage{amsthm}
\IEEEoverridecommandlockouts
\usepackage{times}
\usepackage[scriptsize]{subfigure}
\usepackage{paralist}
\usepackage{mathrsfs}
\usepackage{graphicx}
\usepackage{tabularx}
\usepackage{url}
\usepackage{mathbbol}
\usepackage{booktabs}
\usepackage{verbatim}
\usepackage[ruled,vlined]{algorithm2e}

\usepackage{bm}    
\usepackage{color}

\usepackage{xcolor}
\usepackage{xspace}
\usepackage{cleveref}
\usepackage{epstopdf}

\defaultleftmargin{2em}{}{}{}

\begin{document}

\title{Improved Methods of Task Assignment and Resource Allocation with Preemption in Edge Computing Systems}

\author{Caroline~Rublein,~\IEEEmembership{Graduate~Student~Member,~IEEE},~Fidan~Mehmeti,~\IEEEmembership{Member,~IEEE}, Mark~Mahon,~\IEEEmembership{Senior~Member,~IEEE} and~Thomas~F.~La~Porta,~\IEEEmembership{Fellow,~IEEE}
\thanks{C. Rublein (clr292@psu.edu), M. Mahon (mpm114@psu.edu), and T. La Porta (tfl12@psu.edu) are with the Department of Computer Science and Engineering, The Pennsylvania State University, USA.}
\thanks{F. Mehmeti (fidan.mehmeti@tum.de) is with the Chair of Communication Networks, Technical University of Munich, Germany.}
\thanks{This paper is an extension of \cite{Rublein2022}, which was presented at ICCCN 2022.}
\thanks{This work has been submitted to the IEEE for possible publication. Copyright may be transferred without notice, after which this version may no longer be accessible.}
}

\maketitle

\begin{abstract}
Edge computing has become a very popular service that enables mobile devices to run complex tasks with the help of network-based computing resources. However, edge clouds are often resource-constrained, which makes resource allocation a challenging issue. In addition, edge cloud servers must make allocation decisions with only limited information available, since the arrival of future client tasks might be impossible to predict, and the states and behavior of neighboring servers might be obscured.
We focus on a distributed resource allocation method in which servers operate independently and do not communicate with each other, but interact with clients (tasks) to make allocation decisions. We follow a two-round bidding approach to assign tasks to edge cloud servers, and servers are allowed to preempt previous tasks to allocate more useful ones. We evaluate the performance of our system using realistic simulations and real-world trace data from a high-performance computing cluster. Results show that our heuristic improves system-wide performance by 20-25\% over previous work when accounting for the time taken by each approach. In this way, an ideal trade-off between performance and speed is achieved.
\end{abstract}

\begin{IEEEkeywords}
Edge cloud computing, Optimization, Bidding, Greedy algorithms. 
\end{IEEEkeywords}

\section{Introduction}
\label{Introduction}
Through edge computing, mobile computing devices can request complex processing services by linking to service providers through wireless links~\cite{Abbas2018}. These services frequently include big data analytics (such as video analytics) and real-time control applications (e.g., for smart grid).  Data from these applications may require a significant amount of computing power, memory and network bandwidth for its analysis~\cite{Abbas2018}. Edge cloud resources must be allocated efficiently to process these jobs because edge resources are somewhat limited and the characteristics of arriving jobs can be unpredictable. 

A majority of previous work has focused on centralized systems for resource allocation in which a central entity with complete knowledge of network resources and incoming requests makes resource allocation decisions~\cite{KUMAR2017234}. These decisions are often efficient, but there are several drawbacks to relying on this central entity. These drawbacks include scalability and the need to maintain an accurate status of resource availability, failures, and maintenance for all servers in real time. In addition, if servers are owned by different service providers, they may not want to share their state with a central source~\cite{Rublein2021}. Thus, to circumvent these issues, we utilize a distributed approach for allocating and assigning edge resources to submitted jobs.

The process of accepting jobs (i.e., assigning them to different servers) considers the value (utility) of the submitted jobs, their resource requirements, and the state of the servers \textcolor{black}{(in terms of their occupancy)}.  Resource requirements can include required CPU cycles based on the job and its deadline, network bandwidth for uploading data and downloading results, and memory.

Preemption is the process of exchanging a currently-running job for a newly-arriving job that is deemed more valuable. In an online resource allocation system, preemption can be used to increase utility achieved at the expense of the preempted jobs. Much existing work on distributed resource allocation~\cite{Rublein2021}, \cite{KUMAR2017234}, \cite{Grandl2014}, \cite{tasiopoulos2018edge} has focused on non-preemptive systems. In non-preemptive systems, a task will always run to completion once it is allocated. This can result in some high-value or high-importance tasks being blocked from allocation by other less valuable tasks that arrived beforehand. Determining how often, and under what circumstances, to preempt particular job(s) is a delicate balancing act. If preemption happens too frequently, servers may waste their limited resources on jobs that do not finish. If too few jobs are preempted, then the system performance will match that of a non-preemptive one, thus wasting any extra time spent on preemption decisions.

In this work, we consider an online distributed resource allocation system that allows preemption. Users submit job requests with \textcolor{black}{their} resource requirements, deadline, and utility.  Servers choose which jobs to serve based on their internal state and the qualities of the incoming requests, with the ultimate goal of maximizing the overall utility of served jobs \textcolor{black}{across the system}. We consider resources such as network bandwidth and processor usage to be elastic in that the allocations for these resources may change over time as long as the jobs are completed by their deadline.

One important aspect of a resource allocation algorithm is how long it takes to execute.  If the allocation algorithm takes too long, jobs have less time to complete before their deadlines making it difficult for the system to accept jobs with short deadlines. Our system uses an online, light-weight distributed resource allocation algorithm based on simple bidding.  While more sophisticated approaches exist, they require more time to execute, thus delaying the start of submitted jobs and making it more difficult for jobs with short deadlines to be accepted and complete their processing.  

In this paper, we make the following contributions:
\begin{itemize}
  \item We formally define an optimization problem for an online, elastic resource allocation system that allows preemption for a pipeline processing paradigm.
  \item We describe a greedy heuristic to maintain the scalability of the system and the quality of preemption decisions at scale.
  \item We test the formulation, our algorithms, and previous work using a real-world workload trace from a high-performance computing cluster and show that our approach outperforms state-of-the-art techniques.
  \item We show that our greedy heuristic improves performance by $20$-$25$\% over another recent heuristic for jobs with short deadlines.
\end{itemize}

The remainder of this paper is organized as follows. We discuss some related work in Section~\ref{Related Work}. Then, in Section~\ref{System Overview}, we describe the system overall. This description is followed by the optimization problem formulation in Section~\ref{Formulation}. In Section~\ref{Heuristic}, we present our approach, and we evaluate this approach in Section~\ref{PerformanceEvaluation} using extensive realistic simulations. Finally, Section~\ref{Conclusion} concludes the paper.

\section{Related Work}
\label{Related Work}

Within the realm of edge computing, a wide range of resource allocation approaches exist, such as: application placement, resource scheduling, task offloading, load balancing, resource allocation, and resource provisioning~\cite{ghobaei2019resource}, \cite{Fang2020}, \textcolor{black}{\cite{Wang2015}, \cite{Tan2017}, \cite{Atakan2018}, \cite{Urgaonkar2015}}.

The concept of multi-dimensional bin packing has been proposed in~\cite{mdmbinpack}, whereas basic auction-based resource allocation mechanisms in cloud computing, and auction variations, are described in \cite{wangbasic}. Additional auction mechanisms that can be applied to cloud computing are presented in~\cite{KUMAR2017234}. In general, these focus more on theoretical performance, omitting the idiosyncrasies of real-world applications.

While in most of the proposed auction mechanisms truthfulness is a concept that is user-related, a truthful multi-unit double auction mechanism was proposed in~\cite{Segal-HaleviHA18} that enforces both users and servers to act truthfully.

Edge-MAP~\cite{tasiopoulos2018edge} proposes a client-to-cloud model for tasks with extremely short deadlines (lower than $100$\:ms). Using a Vickrey-English-Dutch (VED) auction~\cite{ANDERSSON2013}, the system achieves a unique minimum competitive equilibrium price. Due to this property, the system is scalable and adjustable to network topologies that change, but it does rely on a central auctioneer for a set of servers.

An alternative market-based framework by Nguyen \emph{et al.} allows for resources to be efficiently allocated by edge nodes that are dispersed~\cite{Nguyen2018}, where market equilibrium is achieved by finding an optimal allocation of resource bundles to services such that the task budget is not violated.

Ant Colony Optimization (ACO) algorithms have been proposed as an alternative method of resource allocation~\cite{CHENG2019}. Using the swarm intelligence provided by an ACO-based method, a single \textcolor{black}{multi-access edge computing} server can decide how to serve each task, with a joint emphasis on minimizing latency and energy consumption. However, in~\cite{CHENG2019} it is assumed that all submitted tasks are of equal value, which cannot be applied to many cases. On the other hand, we assume that different jobs have different values and deadlines.

A convex iterative algorithm to optimize energy consumption and task offloading simultaneously was proposed in~\cite{ZHAO2021}. However, the authors acknowledge that its real-time implementation is not yet possible.

The Dynamic Priority Task Scheduling Algorithm (DPTSA), proposed in~\cite{LIAO2020}, accounts for dynamically-changing task urgencies (similar to deadlines) in conjunction with task priority levels. However, the scale is limited to only one server, which makes the work less general in its scope.

Inter-user task dependencies can also be considered; that is, two user tasks are dependent on each other such that task data must be transferred between mobile devices. In exploring this concept, Yan \emph{et al.}~\cite{YAN2020} focus on the interactions of the mobile devices rather than the resource allocation on the server side.

Other scheduling systems, such as those for multi-resource clusters \cite{Grandl2014}, also aim to maximize task completion while dealing with online arrivals. One such system, Tetris~\cite{Grandl2014}, aims for a balance between fairness and performance, but relies on a central algorithm to distribute tasks between machines. 

A distributed iterative auction using a modified Vickrey mechanism
is proposed in~\cite{ZHANG2017}. Users cannot bid higher than their own utility, and service providers use a two-dimensional knapsack to decide which users to serve. The resulting multi-round sealed sequential combinatorial auction (MSSCA) shows better performance than its simpler counterparts (Random SCA and One-shot Sealed SCA). Other multi-server resource allocation scenarios may also consider the physical distance between users and base stations~\cite{TRAN2019}. This work also uses a heuristic algorithm to approximate the optimal solution since the optimization is too complex. However, in both of these works, the utility of the tasks is purely defined as the weighted sum of the improvement in task completion time and device energy consumption, leaving no room for tasks to have their own inherent value.

We have thoroughly compared our work to the double knapsack system described in \cite{Rublein2021}, which also analyzed the effects of varying utility disclosure levels and basic dishonesty prevention. Other prior work on this type of double auction bidding system focuses on testing a clustering-based method for scalability and assumes jobs follow a 3-step batch processing paradigm \cite{Rublein2022}.

Zhang \emph{et al.}~\cite{Zhang2023} examine a system with heterogeneous utility functions, namely sigmoid, linear, and all-or-nothing. In this system, servers also seek to maximize their total utility, and preemption is allowed. The main algorithm presented is centralized, but a modification is presented such that it can be used in a distributed fashion. After each server quotes a tentative finish time to each job, each job selects a server based on the earliest quoted time, and this selection is final. This method relies on accurate time estimation to maintain quality of service for users and removes an additional opportunity for servers to make decisions about incoming jobs.

Another work \cite{Meng2019} considers the effect of task dispatching latency (i.e., latency while sending a job to a server) and other communication-related delays. However, tasks are allowed to miss their deadlines after being allocated.

In~\cite{Vaji_Infocomm}, the authors consider the problem of joint service placement and request scheduling that maximizes the expected number of requests served per time slot, given various constraints in terms of the resources. To achieve this, the authors propose a two-time-scale framework, where the service placement is performed on a longer time scale (of frames), whereas the request scheduling is performed on the shorter time scale of slots. The proposed algorithm in~\cite{Vaji_Infocomm} achieves $90$\% of the optimal performance. The main difference in our setup is that we try to maximize the total utility of served jobs, and we also consider the case where the users are not truthful. 

\section{System Overview}
\label{System Overview}
\begin{figure}[t]
\includegraphics[width=8cm]{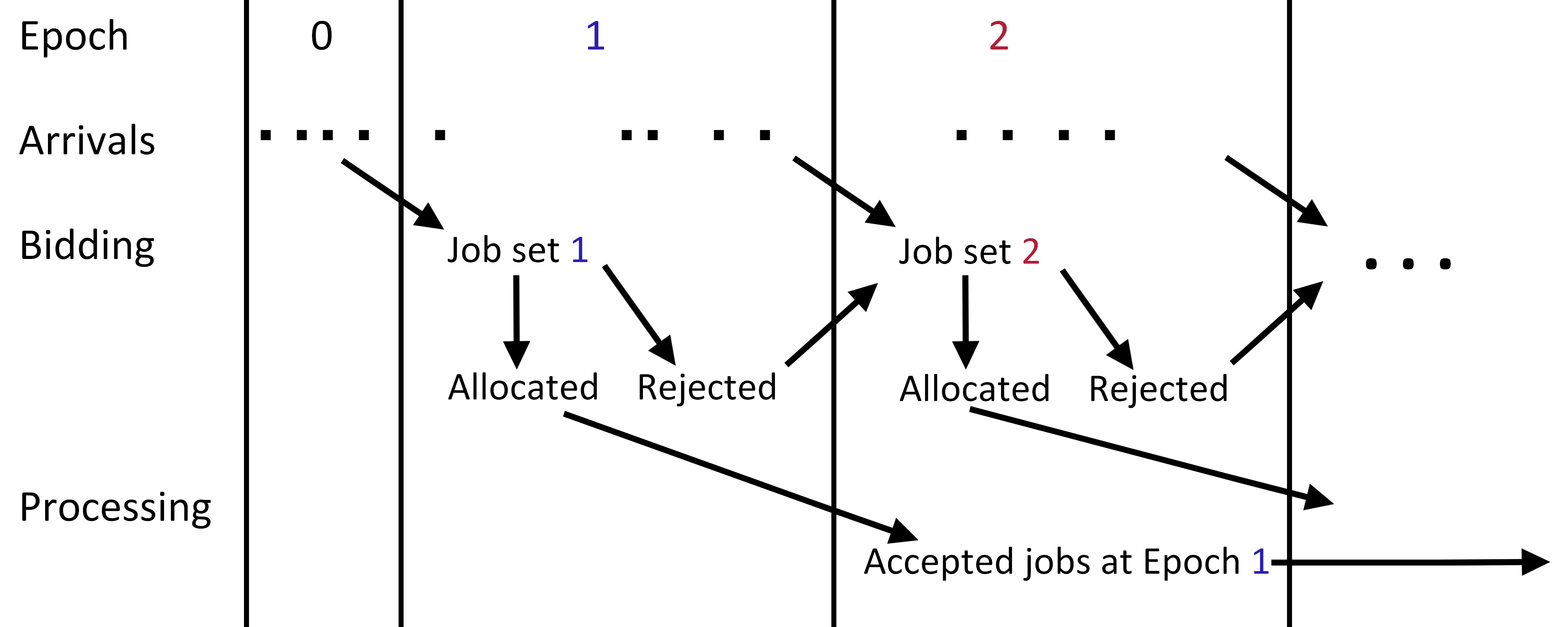}
\caption{The arrival of jobs, along with the bidding and processing procedures.}
\label{fig:jobProcedures}
\end{figure}

The goal of this system is to efficiently allocate processing resources to mobile clients that submit jobs to access edge computing resources over constrained wireless links. Our resource allocation algorithm consists of two phases: \textit{bidding} and \textit{processing}. In the bidding phase, a two-round double auction takes place in which clients and servers each decide their own best course of action with respect to assigning jobs to a server. As soon as the bidding phase for a particular set of jobs is complete, their processing phase begins. This process is illustrated in Fig.~\ref{fig:jobProcedures}, reproduced here from \cite{Rublein2022} for convenience. For example, in bid epoch 2, the jobs accepted in epoch 1 begin their processing phase. Likewise, the jobs arriving in epoch 2 will begin their bidding phase in epoch 3 and their processing phase in epoch 4.

Each bidding phase follows an internal two-round structure. In Round 1, clients submit job requests to all available servers. These requests include information about the job's required resources, as well as the job's stated utility (\textcolor{black}{how much it is worth}). In response to these requests, servers set bid prices for each submitted job.  In Round 2, clients choose the server that offered them the cheapest price and request that server to process their job. The servers then decide which returning job requests are allocated resources. If a client's job did not make the cut in the auction, they may choose to resubmit their job in the next bidding phase.

How Round 1 prices are set has a major impact on the performance of the system. Servers have no guarantee that any job that submits a request in the first round will return for a second round because it does not know the prices bid by other servers.  If a server sets its prices too high, not enough jobs will return for Round 2 and the server will be underutilized. Conversely, if Round 1 prices are too low, users lose information that would help them select a server more suited for their job(s), and many more jobs than can be accommodated by a server may return in Round 2. This can lead to congestion at one server while others are starved depending on the server size distribution across different resources. In addition, if many jobs return to one server in Round 2 while others are starved, some jobs will be rejected that could have fit on other servers.  Once jobs are accepted in Round 2, processing begins.

There are two main paradigms that can be followed when considering the internal structure of the processing phase: \textit{batch} and \textit{pipeline}. In batch processing, jobs progress through 3 distinct upload, processing, and download phases. When a job is accepted, it first uploads all of its required \textcolor{black}{content (data)} to the server. Then, no bandwidth will be used during the processing phase while processing resources are occupied. When the job begins downloading results to the user, the processing resources it used will be freed. Thus, processing resources and bandwidth will not be used simultaneously for a single job. By contrast, in pipeline processing, all resources may be used simultaneously. For example, a job that uploads 30\% of its required data to the server may then have that content processed and even downloaded while it continues uploading.

An additional mechanism that can be included in Round 2 of the bidding phase is \textit{preemption}. In this case a job returning for Round 2 in the bidding phase may preempt a job that has been executing from a previous round.  However, deciding when to preempt is not straightforward, as measuring one job's value/costs against another job's depends on several factors. Namely, it is difficult to determine the exact value relationship between stated utility, time, and resources such that each job may be fairly compared to every other job. In addition, if utility is only obtained when a job completes, then we must consider how much time remains for that utility to be accrued when deciding if it should be preempted.   

Using preemption further complicates the setting of Round 1 prices. If a job that arrives in Round 1 can fit on a server without preempting any existing jobs it can be given a low price.  However, some jobs may be good candidates to preempt existing jobs, perhaps because they have high utility and do not require a great deal of resources.  In this case a server may be tempted to offer this type of job a low price as well so it returns in Round 2.  The complication is that if the job will fit on a different server without preemption then the system will achieve higher utility if that job returns to the server where it does not preempt any jobs so more jobs in the system can complete.  Therefore when trying to attract jobs that are good candidates to preempt, care must be taken to not set prices so low so as to attract jobs that could fit on other servers without preemption.

\section{Optimization Formulation}
\label{Formulation}

In this section, we provide an optimization formulation that captures a version of the system described in Section~\ref{System Overview} that supports pipeline processing and allows preemption, with jobs arriving over time. \textcolor{black}{The objective is to maximize the served utility over the entire system.} 

This optimization formulation describes a system in which a centralized entity has complete knowledge of all server and job specifications, including those of future job arrivals; thus this entity acts like an \emph{oracle} for the system.

The system consists of $|\mathcal{I}|$ servers, and there are a total of $|\mathcal{J}|$ jobs arriving in the system over the considered time horizon. The system is time slotted, and we consider a finite time horizon of $|\mathcal{N}|$ slots\footnote{We will use the terms \emph{slot} and \emph{timestep} interchangeably in this paper. Note that the slot does not refer to the slot duration in a cellular network.}. The arrival time of task $j$ is $a_j$; the deadline by which it must be finished after the arrival is $d_j$.\footnote{To consider the problem more generically (for any slot duration), these quantities are expressed in terms of slots. E.g., task $j$ arrives in the $5$th slot, and has a deadline of $7$ slots.} The utility obtained if task $j$ is served within the deadline is $U_j$. Task $j$ has a storage requirement of $s_j$ (expressed in MB), and a total computational requirement of $K_j$ (expressed in MFlops).
The total storage capacity of server $i$ is $S_i$ (expressed in GB), whereas its total computational capacity is $C_i$ units per slot (expressed in \textcolor{black}{MFlops/s}). The parameters $B_{u,i}$ and $B_{d,i}$ represent the total upload and download capacity of server $i$, respectively. As we assume that the slot duration is fixed, we can express the last two parameters in terms of the data size that can be sent in the downlink/uplink in a slot. Namely, $B_{u,i}$ ($B_{d,i}$) would be equal to the product of the capacity in the uplink (downlink) and the slot duration. 

There are multiple decision variables. The first four are related to assigning resources and their consumption. These are: (i) $x_{i,j}$, whose value is $1$ if task $j$ is assigned to server $i$, and $0$ otherwise; (ii) $\sigma_j(n)$, which denotes the amount of data that are uploaded on the chosen server for a given task at a given slot; (iii) $\kappa_j(n)$ is the amount of computation resources reserved on slot $n$ to run task $j$ at the corresponding server; (iv) $\sigma^{'}_{j}(n)$ denotes the amount of data, in terms of the obtained results, that are sent back to the user after task $j$ has been successfully executed at one of the servers.\footnote{Note that the decision variables $\sigma_j(n)$ and $\sigma_j^{'}(n)$ are directly proportional to the uplink and downlink rate at slot $n$, respectively, i.e., they are equal to the product of the corresponding bandwidth component and the slot duration.}

Other decision variables related to preemption are defined as follows.
The parameter $\tau_j$ is the preemption-related parameter. Its value is $0$ if the task is preempted, and $1$ if the task is run to the end. The accruing policy is all-or-nothing, i.e., if the task is preempted at any point before being completely executed, the earned utility is $0$. Note that the task can be preempted during any phase: upload, processing, or download.    
The decision variable $d_{j,t}$ denotes the number of slots the task spends in processing before it is preempted. If $d_{j,t}=d_{j,d}$, the job is not preempted.

There are an additional three variables that reflect different stages of job processing: (i) $d_{j,u}$ is the number of slots across which the upload of data to the server spans; (ii) $d_{j,p}$ denotes the number of slots across which the task is being executed on the server; and (iii) $d_{j,d}$ is the number of slots that it takes for the server to send back the results of the processed task to the user.

The formulation of the optimization problem is as follows:
\vspace{-12pt}

\footnotesize
\begin{align}
\max & \sum_{i=1}^{|\mathcal{I}|}\sum_{j=1}^{|\mathcal{J}|}U_j\tau_jx_{i,j} \label{eq:obj}\\
\mbox{s.t.} 
& \sum_{l_j=1}^{|\mathcal{N}|}\sigma_j(l_j)\leq s_jx_{i,j},\quad \forall i\in\mathcal{I}, \forall j\in\mathcal{J}, \label{eq:sigmaj_req}\\
& \tau_j\left(\sum_{l_j=1}^{|\mathcal{N}|}\sigma_j(l_j)-s_jx_{i,j}\right)=0,\quad \forall i\in\mathcal{I}, \forall j\in\mathcal{J}, \label{eq:tau_sigmaj}\\
& \sum_{l_j=1}^{|\mathcal{N}|}\kappa_j(l_j)\leq K_jx_{i,j},\quad \forall i\in\mathcal{I}, \forall j\in\mathcal{J}, \label{eq:kappaj_req}\\
& \tau_j\left(\sum_{l_j=1}^{|\mathcal{N}|}\kappa_j(l_j)-K_jx_{i,j}\right)=0,\quad \forall i\in\mathcal{I}, \forall j\in\mathcal{J}, \label{eq:tau_kappaj}\\
& \sum_{l_j=1}^{|\mathcal{N}|}\sigma^{'}_j(l_j)\leq s^{'}_jx_{i,j},\quad \forall i\in\mathcal{I}, \forall j\in\mathcal{J}, \label{eq:lsigmaj_req}\\
& \tau_j\left(\sum_{l_j=1}^{|\mathcal{N}|}\sigma^{'}_j(l_j)-s^{'}_j\right)=0,\quad \forall j\in\mathcal{J}, \label{eq:tau_lsigmaj}\\
& \frac{\sum_{l_j=1}^{|\mathcal{N}|}\sigma^{'}_{j}(l_j)}{s^{'}_{j}}<1+\tau_j, \quad \forall j\in\mathcal{J}, \label{eq:results_tau}\\
& \sum_{l_j=1}^{n}\kappa_j(l_j)\leq\frac{\sum_{l_j=1}^{n}\sigma_j(l_j)}{s_j}K_j, \quad \forall n\in\mathcal{N}, \forall j\in\mathcal{J}, \label{eq:proc_limit}\\
& \sum_{l_j=1}^{n}\sigma^{'}_j(l_j)\leq\frac{\sum_{l_j=1}^{n}\kappa_j(l_j)}{K_j}s^{'}_j, \quad \forall n\in\mathcal{N}, \forall j\in\mathcal{J}, \label{eq:down_limit}\\
& d_{j,u}\leq d_{j,p}\leq d_{j,d}\leq d_j, \quad \forall j\in\mathcal{J}, \label{eq:int_deadlines}\\
& d_{j,u}\geq 1, \quad \forall j\in\mathcal{J}, \label{eq:dju}\\
& d_{j,p}\geq 1, \quad \forall j\in\mathcal{J}, \label{eq:djp}\\
& d_{j,d}\geq 1, \quad \forall j\in\mathcal{J}, \label{eq:djd}\\
& \sum_{j=1}^{|\mathcal{J}|}\sum_{l_j=1}^{n}\sigma_j(l_j)x_{i,j}\theta_{j}(n)\leq S_i, \quad \forall i\in\mathcal{I}, \forall n\in\mathcal{N}, \label{eq:svr_stor}\\
& \sum_{j=1}^{|\mathcal{J}|}\kappa_j(n)x_{i,j}\leq C_i, \quad \forall i\in\mathcal{I}, \forall n\in\mathcal{N}, \label{eq:svr_comp}\\
& \sum_{j=1}^{|\mathcal{J}|}\sigma_j(n)x_{i,j}\leq B_{u,i}, \quad \forall i\in\mathcal{I}, \forall n\in\mathcal{N}, \label{eq:svr_up}\\
& \sum_{j=1}^{|\mathcal{J}|}\sigma^{'}_j(n)x_{i,j}\leq B_{d,i}, \quad \forall i\in\mathcal{I}, \forall n\in\mathcal{N}, \label{eq:svr_down}\\
& \sum_{i=1}^{|\mathcal{I}|}x_{i,j}\leq 1, \quad \forall j\in\mathcal{J}, \label{eq:xij}\\ 
& x_{i,j} \in \{0,1\}, \quad \forall i\in\mathcal{I}, \forall j\in\mathcal{J}, \label{eq:xij_def}\\
& \tau_{j} \in \{0,1\}, \quad  \forall j\in\mathcal{J}, \label{eq:tau_def}\\
& \sigma_j(n)=0, \quad \forall j\in\mathcal{J},  n=\{1,\ldots,a_{j}-1,a_j+\min\{d_{j,u}, d_{j,t}\},\ldots,|\mathcal{N}|\}, \label{eq:sigmaj_eq0}\\
& \sigma_j(n)\geq 0, \quad \forall j\in\mathcal{J}, n=\{a_j,\ldots,a_j+\min\{d_{j,u}, d_{j,t}\}-1\}, \label{eq:sigmaj_geq0}\\
&\begin{aligned}
\kappa_j(n)=0, & \quad \forall j\in\mathcal{J},\\
& n=\{1,\ldots,a_{j},a_j+\min\{d_{j,p}, d_{j,t}\},\ldots,|\mathcal{N}|\}, \\
\end{aligned} \label{eq:kappaj_eq0}\\
&\begin{aligned}
\kappa_j(n)\geq 0, & \quad \forall j\in\mathcal{J},\\
& n=\{a_j+1,\ldots,a_j+\min\{d_{j,p},d_{j,t}\}-1\},\\
\end{aligned} \label{eq:kappaj_geq0}\\
&\begin{aligned}
\sigma^{'}_j(n)=0, & \quad \forall j\in\mathcal{J}, \\
& n=\{1,\ldots,a_{j}+1,a_j+\min\{d_{j,d},d_{j,t}\},\ldots,|\mathcal{N}|\},\\
\end{aligned} \label{eq:lsigmaj_eq0}\\
&\begin{aligned}
\sigma^{'}_j(n)\geq 0, & \quad \forall j\in\mathcal{J},\\
& n=\{a_j+2,\ldots,a_j+\min\{d_{j,d},d_{j,t}\}-1\},\\
\end{aligned} \label{eq:lsigmaj_geq0}\\
& d_{j,t}\in\{1,\ldots,d_{j,d}\}, \quad \forall j\in\mathcal{J}, \label{eq:djt_def}\\
& \tau_j(d_{j,t}-d_{j,d})=0, \quad \forall j\in\mathcal{J}, \label{eq:tau_djt}\\
& \frac{d_{j,t}}{d_{j,d}}<1+\tau_j, \quad \forall j\in\mathcal{J}. \label{eq:djd_tau}
\end{align}
\normalsize

The objective (\ref{eq:obj}) is to maximize the total utility of served jobs in the entire system over time and using an all-or-nothing method of accruing utility.
Constraints (\ref{eq:sigmaj_req}) and (\ref{eq:tau_sigmaj}) are related to the amount of data uploaded to the server over time; if the task is not served, the former would be $0$. If the task is served, but is preempted while running (during any of the three phases), $\tau_j$ in (\ref{eq:tau_sigmaj}) is $0$, still satisfying (\ref{eq:tau_sigmaj}). On the other hand, if the task is served to the end, $\tau_j=0$, forcing the second left-hand side term of (\ref{eq:tau_sigmaj}) to $0$, implying that the complete data for that task have to be stored on the server. 

Constraints (\ref{eq:kappaj_req}) and (\ref{eq:tau_kappaj}) perform the same action in relation to processing. Specifically, only if the task is run and not preempted, it is assumed that the entire task is to be processed. Constraints (\ref{eq:lsigmaj_req}) and (\ref{eq:tau_lsigmaj}) guarantee that if the utility is gathered for executing this task, all its processed data must be received by the user. Also, if the decision is that the task is not to be run, (\ref{eq:lsigmaj_req}) ensures that no data are sent back. 

Constraints (\ref{eq:results_tau})-(\ref{eq:down_limit}) are of a different nature. Constraint (\ref{eq:results_tau}) states that if a task is preempted ($\tau_j = 0$), then the amount of data sent to the user must be less than the total requested amount (otherwise, the job would have completed $100$\%).

\begin{figure}[t]
\includegraphics[width=8cm]{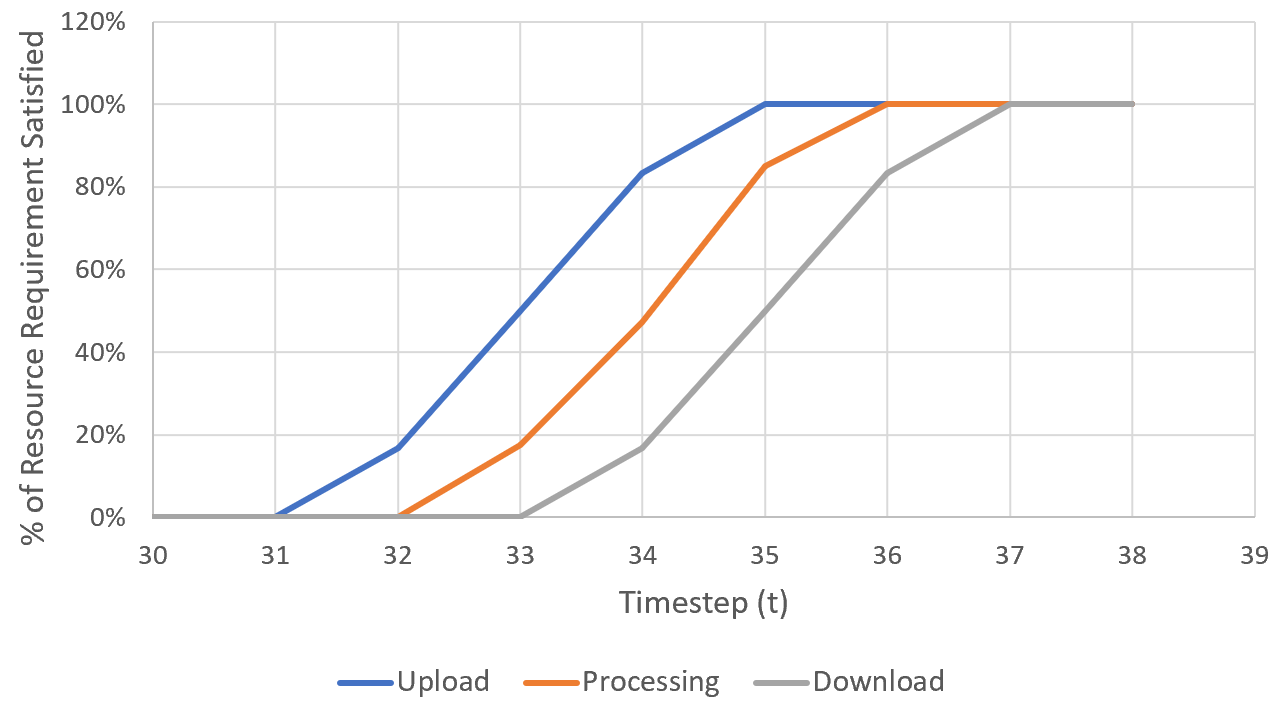}
\caption{Relationship described by constraints (\ref{eq:results_tau})-(\ref{eq:down_limit}), demonstrated using a real-workload job's progression under our KnapsackGreedy heuristic}.
\label{fig:formulation_figure}
\end{figure}

Constraint (\ref{eq:proc_limit}), ensures that the task cannot be processed to a higher extent than there is data stored on the server. This is inherent to the pipeline nature of processing the task, where processing goes on in parallel with storing data. We assume proportionality to that end. E.g., if $60$\% of the data is stored on the server, then no more than $60$\% of the task can be processed. Similarly, (\ref{eq:down_limit}) ensures that the maximum amount of results that are sent back to the user is proportional to the amount of the task that has been processed on the server. In a similar vein as before, if $60$\% of the task is processed by a given slot in the server, at most $60$\% of the result data could have been downloaded to the user by that slot. This relationship between the delay parameters is illustrated in Fig.~\ref{fig:formulation_figure}, which was gathered from a real-workload job's processing phase \textcolor{black}{(see Section~\ref{TraceData} for workload description).}

Constraint (\ref{eq:int_deadlines}) controls the relative order of intermediate deadlines. This differs from the batch formulation~\cite{Rublein2022} in that two phases may complete at the same time. Constraints (\ref{eq:dju})-(\ref{eq:djd}) simply state the fact that each process (uploading, processing, and downloading) needs at least one slot. Constraint (\ref{eq:svr_stor}) captures the finite storage capacity on the server. Note that there is an indicator variable $\theta_{j}$, which denotes the time instants when a part of the server's storage is taken by the task's data. It is defined as

\vspace{-9pt}
\small
\[
\theta_j(n)=
\begin{cases}
1, &
\begin{split}
\text{if} \min\{a_j + l_j| & \kappa_j(a_j + l_j)>0\} \leq n \\
& \leq \max\{a_j + l_j|\kappa_j(a_j + l_j)>0\}
\end{split}
\\
0,& \text{otherwise}.  \hfill \hfill (31)
\end{cases}
\]
\normalsize
It is worth mentioning that this check is made in every slot. If the task is executed and results are sent back, the storage taken by the task on the server is released. 

The limited server resources for computation, upload and download bandwidth are captured by (\ref{eq:svr_comp})-(\ref{eq:svr_down}).
Inequality (\ref{eq:xij}) constrains the assignment of every task to at most one server, while (\ref{eq:xij_def}) expresses the fact that a task is either assigned or not. Constraint (\ref{eq:tau_def}) states that a job is either preempted or not, since we assume all-or-nothing utility for this formulation. The fact that data can be uploaded only during the time the task is active is described by (\ref{eq:sigmaj_eq0}) and (\ref{eq:sigmaj_geq0}). Similarly, the processing activity of the task only while it is active is captured by (\ref{eq:kappaj_eq0}) and (\ref{eq:kappaj_geq0}); whereas the fact that no more results are being sent after the task is completely finished or preempted is described by (\ref{eq:lsigmaj_eq0}) and (\ref{eq:lsigmaj_geq0}). The fact that preemption can occur at any point in time while the task is active is captured by (\ref{eq:djt_def}).    
Finally, the two last constraints (\ref{eq:tau_djt}) and (\ref{eq:djd_tau}) enforce that either the task is preempted, in which case  $d_{j,d}>d_{j,t}$ and $\tau=0$, or it is processed to the end, when $d_{j,d}=d_{j,t}$ and $\tau=1$, i.e., preemption can occur only before the entire job is completed.

There are two main problems with this formulation from a practical standpoint. First, it is highly unrealistic for the system to be able to predict future arrivals exactly, with all the task-related specifications. Second,
even if the system possesses this oracle feature, the optimization problem (\ref{eq:obj})-(\ref{eq:djd_tau}) is a Mixed-Integer Non-Linear Program (MINLP), with the time dimension. Within a single time slot $n$, this is still an MINLP similar to the generalized assignment problem,
which is known to be NP-Hard~\cite{Ozbakir2010}.
In the case of large amount of tasks or a longer time horizon, obtaining a solution to this problem even with a solver, like Gurobi, run on sophisticated servers can take very long (up to several days), which would make the solution useless in dynamic environments.

Therefore, in the next section, we propose a heuristic solution for an online system which provides results close to the optimal.
More on this is then to follow in Section~\ref{PerformanceEvaluation}.  

\section{Heuristic Methods}
\label{Heuristic}
A heuristic is necessary for this problem since it is NP-Hard. The goal of our heuristic is for the system to complete the largest amount of utility possible over all time slots.  We allow preemption and consider an all-or-nothing model of accruing utility. To meet this goal we try to set prices to drive jobs to servers where they can fit without preempting other jobs, and when a job cannot fit on any server, attract them to the server where there is the largest gain if preemption occurs.  

\subsection{Knapsack Greedy Heuristic}

Previous work explored a Double Knapsack method~\cite{Rublein2021}, in which servers use a knapsack algorithm in each round to choose jobs. In that work, the Round 1 knapsack determined which jobs could fit on the server and provided low prices to those jobs; the second round knapsack was used to determine which jobs that returned in Round 2 were admitted to the server. The main drawback of the Double Knapsack method is a long computation time~\cite{Rublein2022}. 

Other work~\cite{Rublein2022} utilized a clustering method in Round 1 while maintaining the Round 2 knapsack. In that work the clustering round grouped jobs into those what were a good fit for a server, and hence received low Round 1 prices, using a faster but less accurate method than a knapsack. Clustering is demonstrably faster, but makes less efficient allocation decisions \cite{Rublein2022}. Thus, in the following subsections we outline a greedy heuristic method for Round 2, and an alternative knapsack method for Round 1. The goal of this is to achieve reasonably good results in a shorter timeframe than Double Knapsack.

\subsubsection{Round 1 Procedure}
In Round 1, a knapsack is run to estimate which requesting jobs will fit on the current residual server space, i.e., do not require preemption. The jobs that fit into this knapsack are given prices that are discounted from their stated utility by $10$\%, i.e., they are given a price that is $10$\% lower than their stated utility. This amount of discount was previously determined to provide the best protection against clients overstating their utility~\cite{Rublein2021}. These jobs are also marked for future reference should they return in Round 2.

It is also beneficial to entice jobs that would provide a significant utility gain if they were to preempt a currently-running job. As discussed above, it is damaging to attract a job that will cause preemption if that job can fit on a different server without preemption. Thus, if a job is not accepted into the knapsack but is still desirable to a server, it will receive a  discount that is capped at less than the discount given to jobs that fit on the server without requiring preemption, thus guaranteeing that if it can fit on a different server without preemption, it will return there for Round 2. The actual discount is determined by two factors.

The first factor is the percentile of this job on the server compared to other current jobs in terms of utility/time\_remaining. For example, if a requesting job is better than $70$\% of the currently-running jobs by that metric, then its percentile will be $0.7$. The first factor is this percentile multiplied by a set constant $c_1$; in our system, $c_1$ is set to $percentile\_weight$.

The second factor is related to the congestion on the server.
It is the sum of the ratio of each requested resource over the server's corresponding residual resource (this expression is illustrated visually in \cite{Rublein2021}). This is also multiplied by a constant factor, $c_2$ \textcolor{black}{(set to $congestion\_weight$)}, to become the congestion factor.

\textcolor{black}{These two factors are then subtracted from the job's stated utility to result in a potential offered Round 1 price. We can set $percentile\_weight$ and $congestion\_weight$ so that the discount given to a job that requires preemption to fit on a server is never as great as the discount given to a job that fits on the server without preemption.  Because the percentile related to utility and the ratio of resources will always be at most $1$, $percentile\_weight$ caps the contribution of the percentile part of the discount, and likewise, $congestion\_weight$ caps the congestion portion of the total discount. Thus, the total discount is capped by $percentile\_weight + congestion\_weight$.  In this paper we set the discount for jobs that fit on a server without preemption (i.e., the 'fit' discount) to $10$\%, and both $percentile\_weight$ and $congestion\_weight$ to be $2.5$\%, thus capping the preemption discount at $5$\%.
Note that the constants $c_1$ and $c_2$ can be set to different values to weight one more than the other, but we found that setting them identically resulted in better load balancing for the servers.}

If a job can never fit on the server under any circumstance (e.g., its required computational resources are greater than the server's total capacity), it will receive a price greater than its utility, which guarantees that it will not return to that server.

Essentially, this process considers several factors that can predict a job's likelihood to succeed in the auction process. Similar to previous work~\cite{Rublein2022}, the knapsack captures jobs that will fit without preemption. For jobs that must preempt to fit, the percentile factor expresses the probability of utility gain, and the congestion factor expresses how easy or difficult the job's space requirements are to accommodate. 

The non-preemptive version of Round 1 is the same since its goal of drawing desirable jobs back to the server remains the same.

\begin{algorithm}
\SetAlgoLined 
  \For {Server s} {
    jobsThatFit = Knapsack(s.residual\_resc, requesting\_jobs);\\
    
    \For{job in requesting\_jobs} {
        \eIf{job in jobsThatFit} {
            price = job.totalUtility * 0.9;\\
        } {
            percentileFactor = $c_1$ * percentile(job, currentJobs);\\
            congestionFactor = $c_2$ * (1-congestion(job, s.residual\_resc));\\
            price = job.totalUtility \textcolor{black}{- (percentileFactor + congestionFactor) * job.totalUtility};\\
        }
    }

  }
 
 \caption{Round 1 Algorithm}
 \label{r1Algorithm}
\end{algorithm}

\subsubsection{Preemptive Round 2 Procedure}
First, all the jobs that return for Round 2 that were marked in Round 1 (based on fitting in the knapsack) are admitted to the server. The remaining returning jobs are sorted by highest-to-lowest utility/time\_remaining (to be evaluated in that order). Each of these jobs is then checked to see if it fits individually on the residual server space. If it does, then it is admitted to the server. If it does not, then a preemption decision must be made. 

When deciding whether or not to admit a job by preempting another, the relative value of the two jobs must be compared against each other. 
A new job must be more valuable than a currently-running one in order for an associated preemption to provide value to the server. In addition, the new job is constrained by the amount of server space consumed by the currently-running one plus any residual resources on the server. Therefore, we allow a new job to be admitted through preemption only if its utility\_deadline is at least $5$\% greater than that of a current job, and it fits in the remaining space (of the current job plus already-available residual space).

\begin{algorithm}
\SetAlgoLined 
  \For {Server s} {
    Admit all autoFit jobs (marked in R1);\\
    sort(returning\_jobs, descending utility/time\_remaining);\\
    sort(s.jobs, ascending utility/time\_remaining);\\
    \For{job in remaining returning\_jobs} {
        \eIf{job fits on s.residual\_resc} {
            Add job to server;\\
        } {
            \For{sJob in s.jobs} {
                \If{job.utility/deadline * $1.05 \geq$ sJob.utility/time\_remaining and job.space $\leq$ (sJob.space + s.residual\_rsc)} {
                    Preempt sJob;\\
                    Add job to server;\\
                }
            }
        }
    }
}

 \caption{Preemptive Round 2 Algorithm}
 \label{r2Algorithm}
\end{algorithm}

\begin{figure}[t]
\includegraphics[width=8cm]{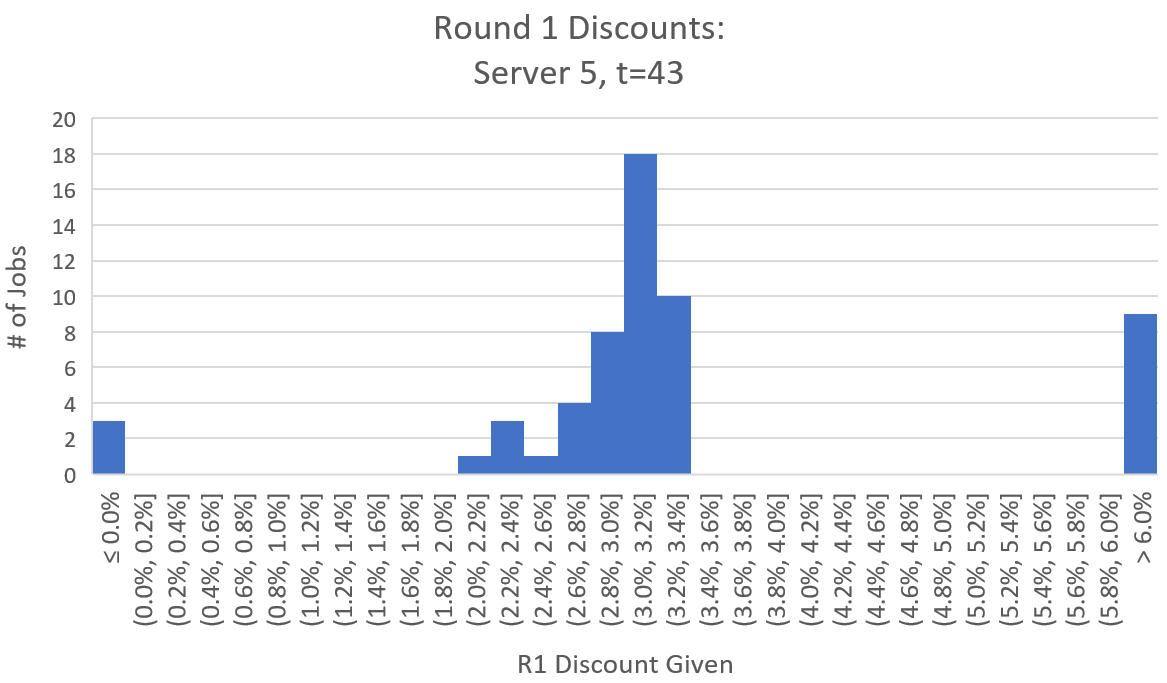}
\caption{Discounts given by Server $5$ to incoming jobs in Round 1 of timestep 43 during a run of the KnapsackGreedy heuristic (pipeline paradigm) using a normally-distributed workload.}
\label{fig:t43_histogram}
\end{figure}

Fig.~\ref{fig:t43_histogram} shows the Round 1 discounts given by a single server utilizing the KnapsackGreedy heuristic. Nine jobs fit in the knapsack, three were rejected because they would not fit were the server to be empty, and the rest were given 'preemption price' discount totals ranging from $2$\% to $3.4$\%. With both $c_1$ and $c_2$ set to $0.025$, the maximum discount possible for a job that does not fit in the knapsack is $5$\%. Experimentally, we found the maximum discount given to prospective preemptive jobs was $4.7$\%. This is because in practice, the congestion discount $c_2$ never reaches its maximum, since every job requires some resources. Parameters $c_1$ and $c_2$ could be set to other constants with similar results; as long as their sum never exceeds $0.1$, the preemption discount is guaranteed to not exceed the `fit' discount.

\begin{figure}[t]
\includegraphics[width=8cm]{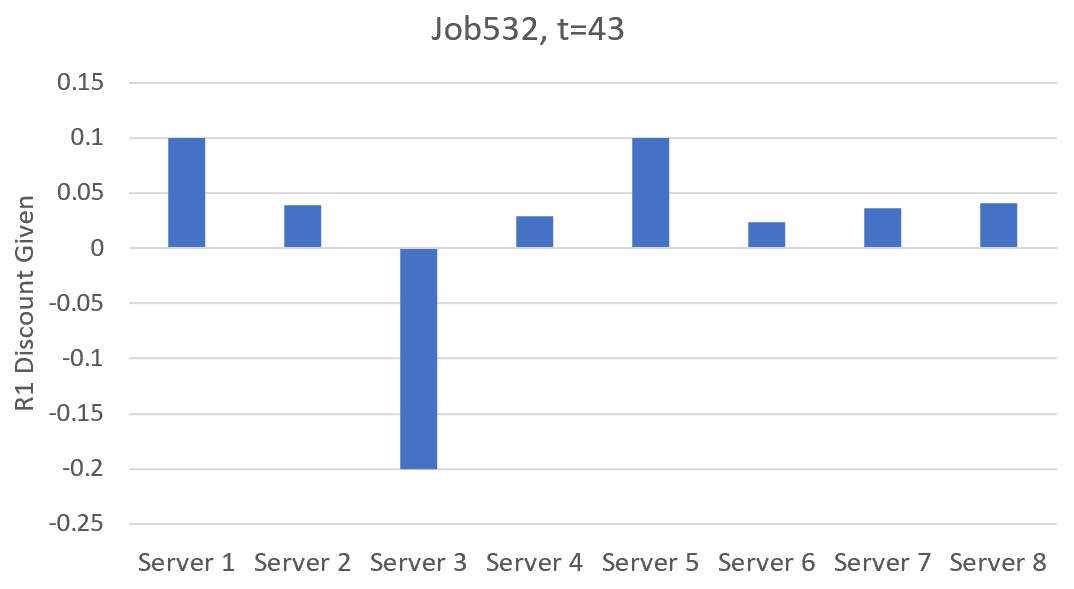}
\caption{Discounts received by job number 532 in Round 1 of timestep 43 during a run of the KnapsackGreedy heuristic (pipeline paradigm) using a normally-distributed workload.}
\label{fig:j532_histogram}
\end{figure}

\begin{figure}[t]
\includegraphics[width=8cm]{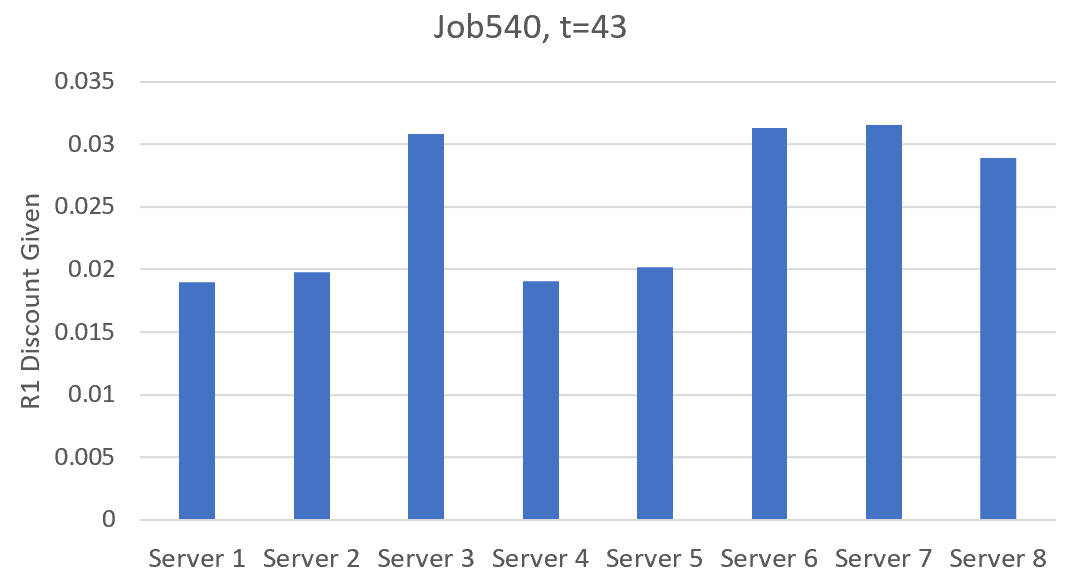}
\caption{Discounts received by job number 540 in Round 1 of timestep 43 during a run of the KnapsackGreedy heuristic (pipeline paradigm) using a normally-distributed workload.}
\label{fig:j540_histogram}
\end{figure}

Figs.~\ref{fig:j532_histogram} and \ref{fig:j540_histogram} show the prices received by two particular jobs from the same run as Fig.~\ref{fig:t43_histogram}. Job 532 fit in the knapsacks of Server 1 and 5, and received preemption prices from every other server except Server 3, which rejected it. Server 5 was randomly chosen from the two fit prices, and Job 532 was admitted there during the first step of Round 2. On the other hand, job 540 (Fig.~\ref{fig:j540_histogram}) received preemptive prices from every server; that is, it could fit on any of them, but only by preempting another job. It chose Server 7 since its price was the lowest, and was able to successfully preempt another job, which had $40$\% less utility/time\_remaining. If it had gone to any of the servers that offered lower preemptive prices ($1$, $2$, $4$, or $5$), it would have been rejected in Round 2. Thus, we show that our heuristic is effective at steering jobs toward the server(s) most likely to accommodate them, normally or preemptively.

\subsubsection{Complexity Analysis}
The Round 1 knapsack runs in $O(n^g)$ time~\cite{Rublein2021}, where $n$ is the number of jobs in the pool, and $g$ is the number of generations (constant $\approx$30)\footnote{Fewer generations are necessary here since the knapsack does not make any permanent decisions about the jobs.} used in the off-the-shelf genetic algorithm knapsack implementation \textcolor{black}{\cite{pyeasyga2016}}.

Round 2 runs in $O(n_2m)$ time, where $n_2$ is the number of returning jobs and $m$ is the number of currently-running jobs on the server. In our simulations, typically $m << n_2 < n$, so its performance is near-linear. Since Round 1 also completes in polynomial time, the overall complexity is $O(n^g + n_2*m) = O(n^g)$.

\subsection{Double Knapsack with Preemption}
\label{DKP_Preemption}
In this section we add preemption to the Double Knapsack algorithm; only Round 2 must be changed. In Round 2, a knapsack is run on each server's total capacity, using both currently-running jobs and jobs returning from Round 1. Jobs that fit in the knapsack are given a 'score' of $1000+ utility/time\_remaining$, and those that do not are given a score of $1+ utility/time\_remaining$. Then, individual jobs are checked for fit in descending score order. Through this method, the jobs that fit in the knapsack are given first priority, and jobs with higher utility/time\_remaining are given secondary priority.

Using this method, any number of jobs may be preempted. The only advantage enjoyed by currently running jobs is that they are closer to their deadline, so their utility/time\_remaining will tend to be better than newly arriving jobs with similar original deadlines.

\section{Performance Evaluation}
\label{PerformanceEvaluation}
In this section, we evaluate the performance of the KnapsackGreedy heuristic for both pipeline and batch paradigms for job processing.

\subsection{Pipeline Paradigm}
We first compare the performance of KnapsackGreedy (with and without preemption) to Double Knapsack (with and without preemption) for the pipeline paradigm.

\subsubsection{Optimal}
\label{OptimalResults}

A state-of-the-art mathematical optimization solver (Gurobi)~\cite{gurobi} was used to solve the optimization problem described in Section~\ref{Formulation}. This optimal system has perfect knowledge of the entire scenario, including future job arrivals, and can assign resources with high precision. While this is not possible in an online system (our application), it does provide an upper bound on the utility we can achieve for a given scenario.

Since the time taken to run the solver on a problem exponentially increases as the problem size increases, we used a relatively small scenario to compare our heuristics to the optimal. We first attempted to obtain the optimum in the solver for the scenario described in \cite{Rublein2022}, which uses $4$ servers and $25$ jobs that arrive over $4$ timesteps. \textcolor{black}{However, it did not complete for over $10$ days, so we could not be $100$\% confident that the solution it found was the global optimal.}
Therefore, we compare the systems to the optimal using a setting that allowed the optimal to complete in approximately $5.5$ hours. A total of $18$ jobs arrive over $3$ timesteps and are allowed to run to completion, using the same job and server distributions as in~\cite{Rublein2022}. Double Knapsack with and without preemption both finished $10$ jobs ($59$\% of the optimal $17$), whereas KnapsackGreedy with and without preemption finished $9$ and $8$ jobs ($53$\% and $47$\% of optimal), respectively.

When an optimal solution is calculated in this way, jobs can be arranged with very high efficiency because the amount of each resource in each timestep for each job can be meticulously set.
For example, the calculated optimal could give Job A near-zero resources in the beginning of its processing time, then allocate all its resources close to its deadline. Meanwhile, Job B could use the majority of the space while Job A is near-zero, and then have its resources tapered off dramatically when Job A needs them. In contrast to this method, our heuristic requires a certain minimum amount of resources to be given to each job per timestep in order to guarantee completion upon allocation; these minimum resources are used to judge jobs against each other during the auction process. Because of this, there may be cases (such as this 18-job scenario) where jobs that might fit otherwise will not fit under the minimum resource requirement.

If the $18$ jobs described above are decreased to $10$, the Double Knapsack algorithms complete all $10$ ($100$\% of optimal) in $\approx$$15$s, KnapsackGreedy with preemption finishes $9$ in $\approx$$11$s, and KnapsackGreedy without preemption only allocates/finishes $8$ jobs in $\approx$$10$s.

\subsubsection{Simulation Setup} In the following sections, we evaluate the performance of KnapsackGreedy and Double Knapsack (both with and without preemption) under the pipeline paradigm. The primary workload for this analysis is normally-distributed according to the distributions shown in Table \ref{fig:normaldist_table}. Eight servers were used, and the arrival rate of jobs was set to $\mu=14$, $\sigma=4$ jobs per slot.

\begin{table}[ht]
\caption{Normally distributed variables for servers and jobs}
\label{fig:normaldist_table}
\centering
 \begin{tabular}{|c|c|c|} 
 \hline
 Resource & $\mu$ & $\sigma$ \\ [0.5ex]
 \hline\hline
 Storage $S_i$ (MB) & 540 & 30\\
 \hline
 Computation $C_i$ (MFlops/s) & 80 & 20\\
 \hline
 Upload Bandwidth $B_i$ (MB/s) & 120 & 30\\
 \hline
 Download Bandwidth $B_i$ (MB/s) & 120 & 30\\
 \hline
 Storage $s_j$ (MB) & 200 & 20\\
 \hline
 Computation $K_j$ (MFlops) & 100 & 20\\
 \hline
 Upload Bandwidth $b_{u,j}$ (MB/s) & 80 & 10\\
 \hline
 Download Bandwidth $b_{d,j}$ (MB/s) & 80 & 10\\
 \hline
 Deadline $d_j$ (slots) & 10 & 3\\
 \hline
 Utility $U_j$ & 60 & 20\\
 \hline
\end{tabular}
\end{table}

The following subsections will describe the two workloads used to test the pipeline paradigm: Normal Distribution and Normal Distribution Accounting for Auction Time. The other processing paradigm, batch, was tested using three simulated workloads: Normal Distribution, Normal Distribution Accounting for Auction Time, and Bimodal (further described in Table~\ref{fig:bimodaldist_table}).

\subsubsection{Results}
\begin{figure}[t]
\includegraphics[width=8cm]{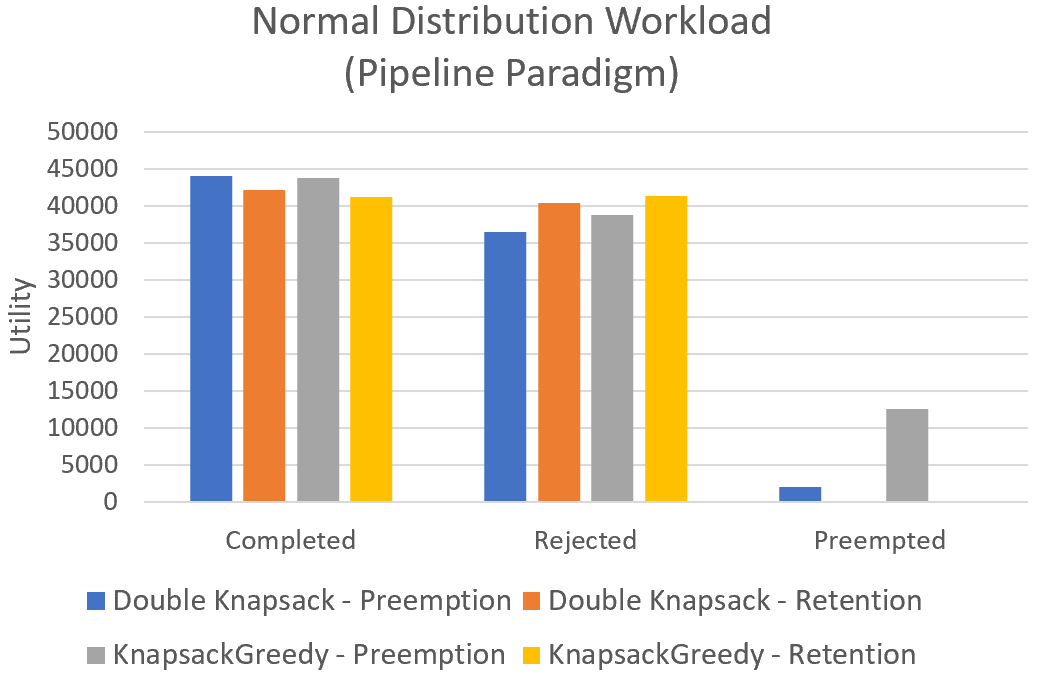}
\caption{Performance comparison of KnapsackGreedy (with and without preemption) to Double Knapsack (with and without preemption) under the pipeline paradigm on a normally-distributed workload.}
\label{fig:pipe_utility_normal}
\end{figure}

\begin{figure}[t]
\includegraphics[width=8cm]{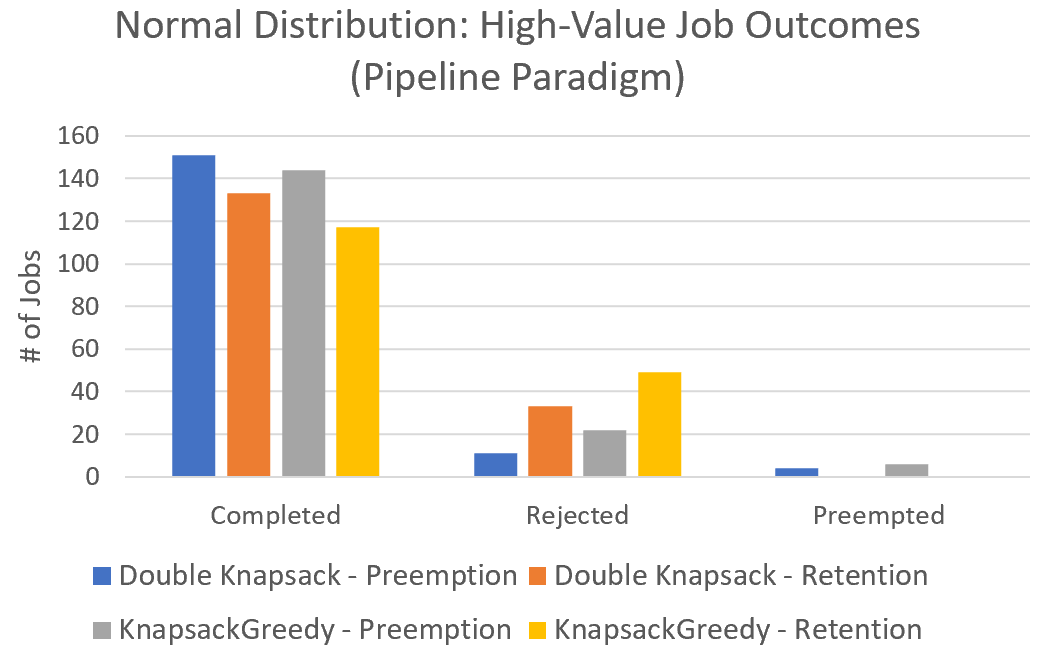}
\caption{Allocation outcomes of high-value jobs for KnapsackGreedy (with and without preemption) and Double Knapsack (with and without preemption) under the pipeline paradigm on a normally-distributed workload.}
\label{fig:pipe_highjobs_normal}
\end{figure}

\begin{figure}[t]
\includegraphics[width=8cm]{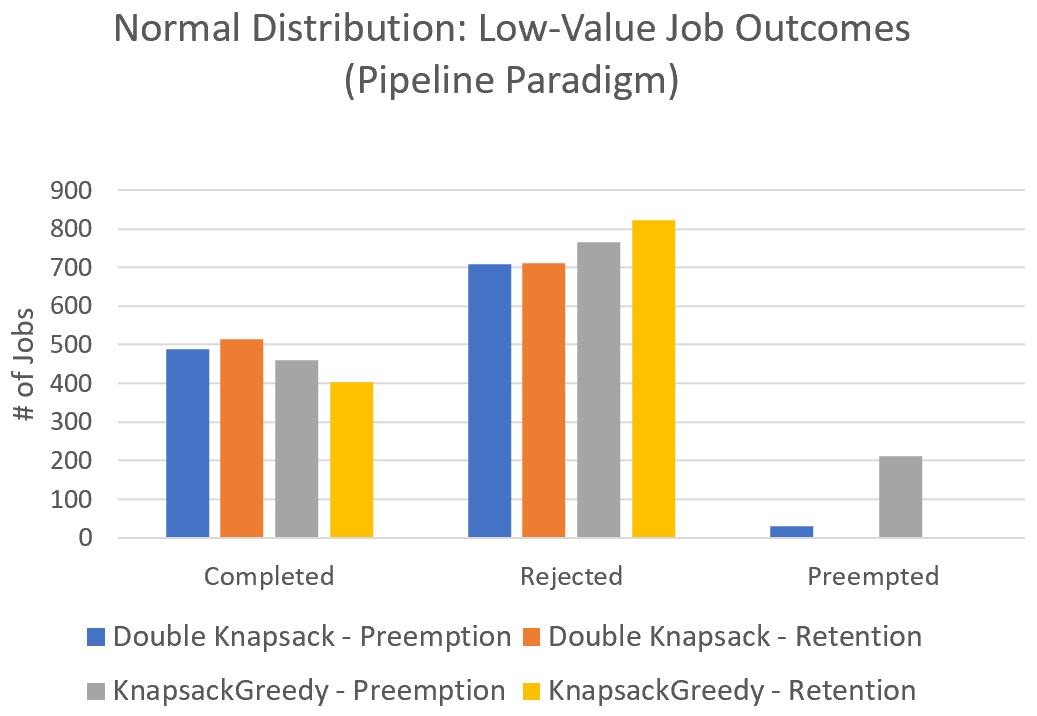}
\caption{Allocation outcomes of low-value jobs for KnapsackGreedy (with and without preemption) and Double Knapsack (with and without preemption) under the pipeline paradigm on a normally-distributed workload.}
\label{fig:pipe_lowjobs_normal}
\end{figure}

The results of the four algorithm variations under the normally-distributed workload are shown in Figs.~\ref{fig:pipe_utility_normal}, \ref{fig:pipe_highjobs_normal}, and \ref{fig:pipe_lowjobs_normal}. Note that the column labeled 'Preempted' depicts the total utility or number of jobs that have been preempted at least once. For both overall utility and number of high-value jobs completed, Double Knapsack (Preemption) $>$ KnapsackGreedy (Preemption) $>$ Double Knapsack (Retention) $>$ KnapsackGreedy (Retention). In general though, the utility difference between the four algorithms is at most $\approx5\%$.

Note that Double Knapsack (Preemption) actually preempts a very small number of jobs, indicating that the base algorithm is very efficient. The KnapsackGreedy preempts approximately four times as much to make up for its less efficient decision making.  This preemption allows the algorithm to achieve performance that is within 2\% of Double Knapsack (Preemption) with a much lower runtime, as discussed next.

\subsubsection{Evaluation Considering Processing Times}
We found that the average per-server auction duration on the simulated workload for Double Knapsack (Preemption), Double Knapsack (Retention), KnapsackGreedy (Preemption), and KnapsackGreedy (Retention) were $\approx5$s, $\approx4$s, $\approx2$s, and $\approx1$s, respectively. If jobs have short deadlines, the length of the bidding process itself takes a significant portion of the time a job has to complete, making it unlikely for jobs with short deadlines to be successful. Therefore, we evaluate the same workload, but account for the differing length of each algorithm's auction process.

\begin{figure}[t]
\includegraphics[width=8cm]{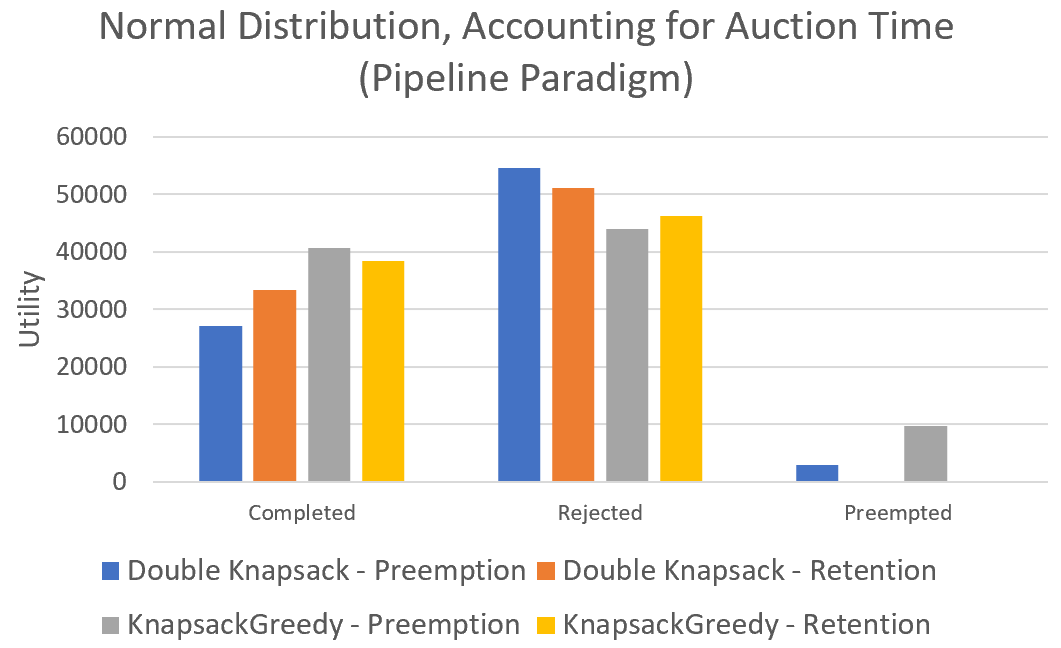}
\caption{Performance comparison of KnapsackGreedy (with and without preemption) to Double Knapsack (with and without preemption) under the pipeline paradigm on a normally-distributed workload with deadlines accounting for auction duration.}
\label{fig:pipe_utility_normal_adjusted}
\end{figure}

The utility achieved by the four algorithms with job deadlines accounting for auction duration is shown in Fig.~\ref{fig:pipe_utility_normal_adjusted}. Since KnapsackGreedy takes less time per auction, jobs are offered more chances to be allocated, and thus more are accepted. Comparing Fig.~\ref{fig:pipe_utility_normal_adjusted} with Fig.~\ref{fig:pipe_utility_normal}, one can see that KnapsackGreedy surpasses the performance of DoubleKnapsack with preemption because the allocation time is significantly shorter. As can be seen in these figures, the KnapsackGreedy (Preemption) algorithm loses almost none of its utility when accounting for the auction time because it executes so quickly. Meanwhile, Double Knapsack (Preemption) loses 34\% of its utility due to its execution time in this case.

\begin{figure}[t]
\includegraphics[width=8cm]{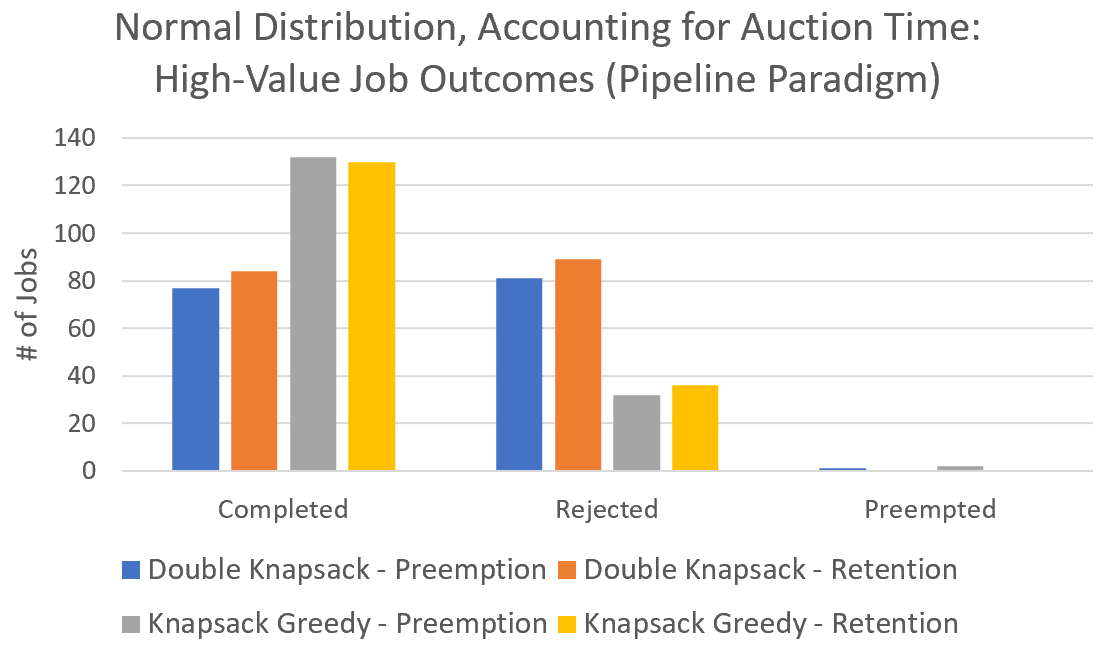}
\caption{Allocation outcomes of high-value jobs for KnapsackGreedy (with and without preemption) and Double Knapsack (with and without preemption) under the pipeline paradigm on a normally-distributed workload with deadlines accounting for auction duration.}
\label{fig:pipe_highjobs_normal_adjusted}
\end{figure}

The difference in performance is particularly significant for high-value job allocations (Fig.~\ref{fig:pipe_highjobs_normal_adjusted}). Both preemptive variations perform slightly better than their non-preemptive counterparts.  Of course if jobs have long deadlines, the difference in the allocation time has less effect.

\subsection{Batch Paradigm}
We also tested the KnapsackGreedy algorithm on jobs following the batch paradigm to cover the cases of jobs that cannot be pipelined.

\subsubsection{Normal Workload}
\begin{figure}[t]
\includegraphics[width=8cm]{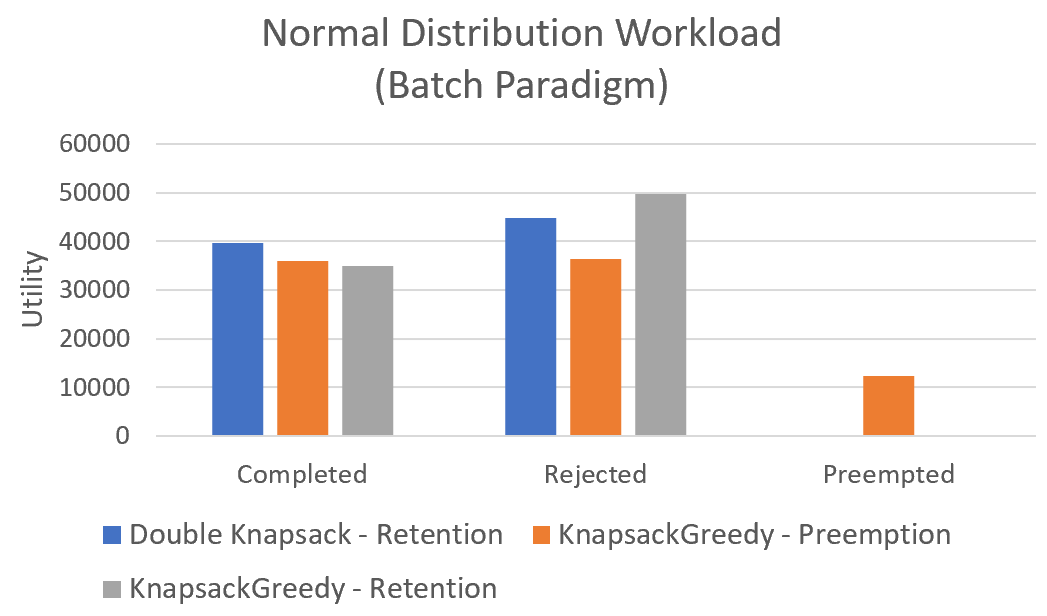}
\caption{Performance comparison of KnapsackGreedy (with and without preemption) to Double Knapsack under the batch paradigm on a normally-distributed workload.}
\label{fig:batch_utility_normal}
\end{figure}

Fig.~\ref{fig:batch_utility_normal} shows the utility achieved by each algorithm under the batch paradigm. The utility of KnapsackGreedy without preemption is very close to that of KnapsackGreedy with preemption. The outcomes for high-value and low-value jobs mirror the overall utility results.

\begin{figure}[t]
\includegraphics[width=8cm]{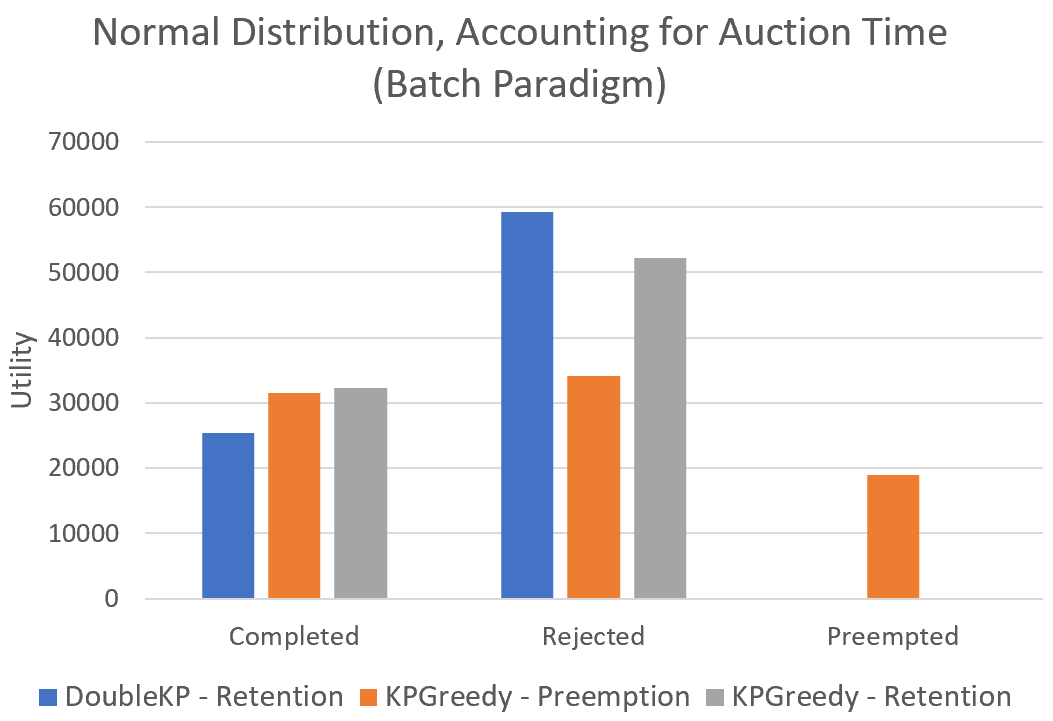}
\caption{Performance comparison of KnapsackGreedy (with and without preemption) to Double Knapsack under the batch paradigm on a normally-distributed workload accounting for auction duration.}
\label{fig:batch_utility_normal_adj}
\end{figure}

The runtimes of the auctions for the batch paradigm are the same as for the pipeline paradigm. To observe the effect of the varied auction durations on system performance, we run the normal workload again, but accounting for each algorithm's auction time. The results are shown in Fig.~\ref{fig:batch_utility_normal_adj}. Similar to the pipeline paradigm, KnapsackGreedy outperforms Double Knapsack because less auction time is taken out of jobs' deadlines.

\subsubsection{Bimodal Workload}
We also simulated a workload with a bimodal distribution for utility. The servers follow the same distribution as described in Table \ref{fig:normaldist_table}, but the job distribution is described in Table \ref{fig:bimodaldist_table}. Exactly $90$\% of jobs are assigned \emph{Utility 1} (low-value), and the remaining $10$\% are assigned \emph{Utility 2} (high-value). Ideally, jobs that belong to the set with the higher mode should be accepted most frequently and preempt jobs from the set with the lower mode.

\begin{table}[ht]
\caption{Bimodal job distribution variables}
\label{fig:bimodaldist_table}
\centering
 \begin{tabular}{|c|c|c|} 
 \hline
 Resource & $\mu$ & $\sigma$ \\ [0.5ex]
 \hline\hline
 Storage $s_j$ (MB) & 160 & 10\\
 \hline
 Computation $K_j$ (MFlops) & 80 & 20\\
 \hline
 Upload Bandwidth $b_{u,j}$ (MB/s) & 70 & 10\\
 \hline
 Download Bandwidth $b_{d,j}$ (MB/s) & 70 & 10\\
 \hline
 Deadline $d_j$ (slots) & 10 & 3\\
 \hline
 Utility 1 $U_{1,j}$ & 40 & 10\\
 \hline
 Utility 2 $U_{1,j}$ & 160 & 20\\
 \hline
\end{tabular}
\end{table}

\begin{figure}[t]
\includegraphics[width=8cm]{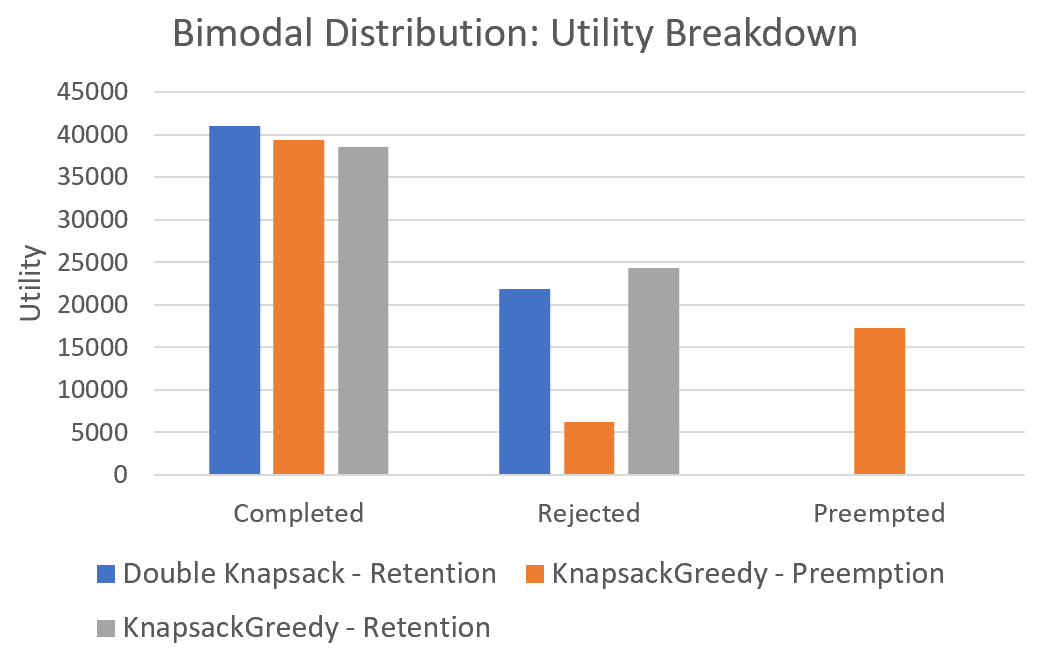}
\caption{Performance comparison of KnapsackGreedy (with and without preemption) to Double Knapsack under the batch paradigm on a workload following a bimodal distribution.}
\label{fig:batch_utility_bimodal}
\end{figure}

The utility achieved on the bimodal workload is shown in Fig. \ref{fig:batch_utility_bimodal}. All three algorithms have somewhat similar performance, but Double Knapsack does slightly better than KnapsackGreedy in terms of overall achieved utility.

\begin{figure}[t]
\includegraphics[width=8cm]{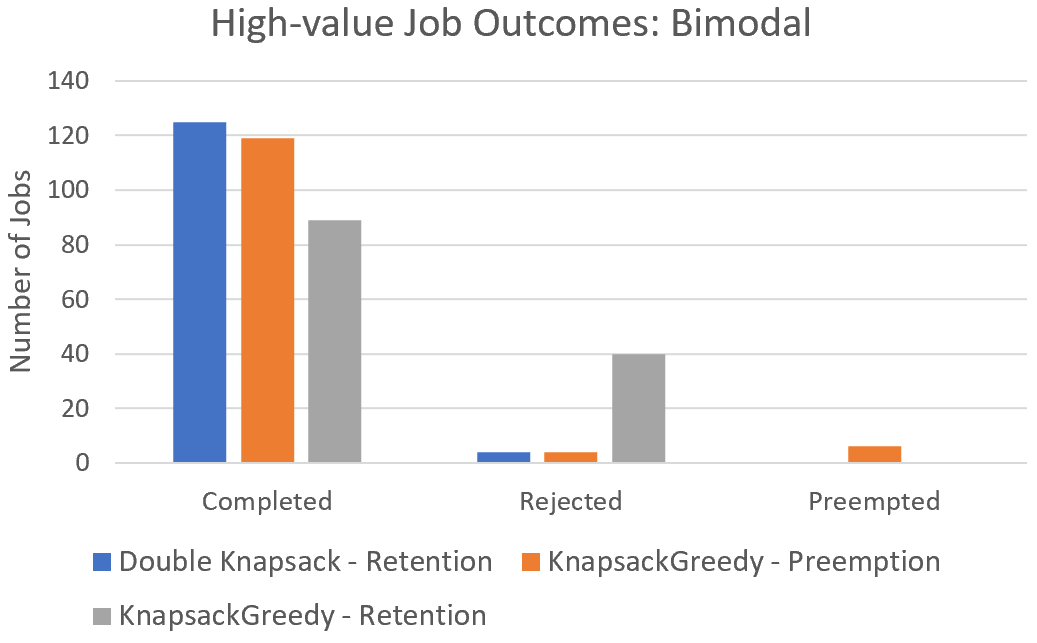}
\caption{Allocation outcomes of high-value jobs for KnapsackGreedy (with and without preemption) and Double Knapsack under the batch paradigm on a workload following a bimodal distribution.}
\label{fig:batch_highjobs_bimodal}
\end{figure}

\begin{figure}[t]
\includegraphics[width=8cm]{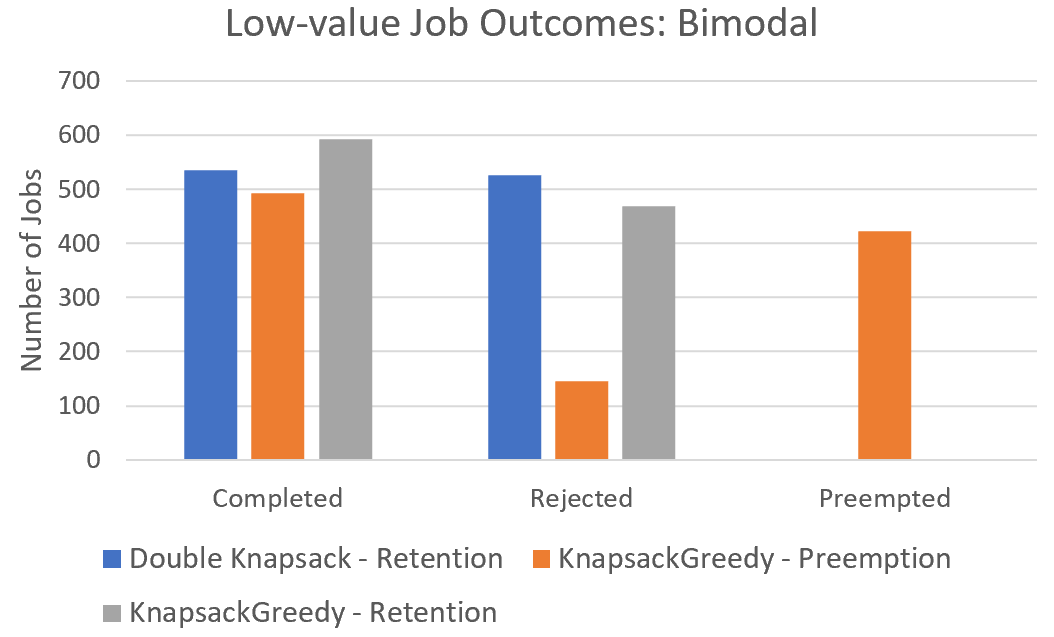}
\caption{Allocation outcomes of low-value jobs for KnapsackGreedy (with and without preemption) and Double Knapsack under the batch paradigm on a workload following a bimodal distribution.}
\label{fig:batch_lowjobs_bimodal}
\end{figure}

Figs.~\ref{fig:batch_highjobs_bimodal} and \ref{fig:batch_lowjobs_bimodal} show the outcomes for high-value and low-value jobs, respectively. There is an interesting difference in that the preemptive version of KnapsackGreedy completes significantly more high-value jobs than KnapsackGreedy (Retention), but completes fewer low-value ones. In addition, a large number of low-value jobs are preempted, whereas only $1$ or $2$ high-value jobs are ever preempted. This points to KnapsackGreedy (Preemption) largely preempting lower-value jobs in order to accept higher-value ones, resulting in its performance more closely resembling Double Knapsack.

\subsubsection{Real Workload Trace}
\label{TraceData}

\begin{figure}[t]
\includegraphics[width=8cm]{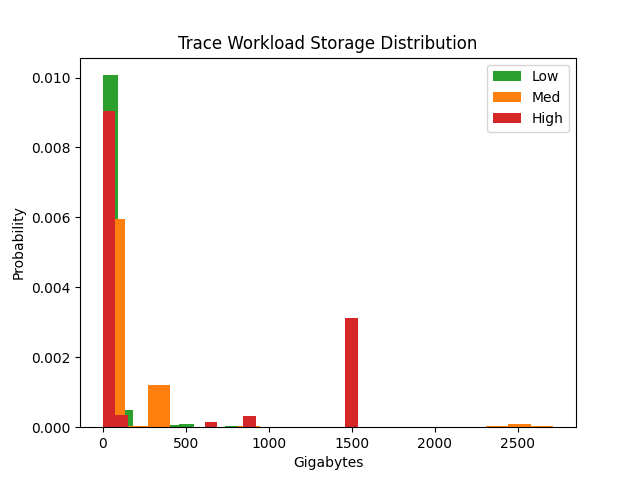}
\caption{Storage distribution from trace data.}
\label{fig:trace_storage}
\vspace{-14pt}
\end{figure}

\begin{figure}[t]
\includegraphics[width=8cm]{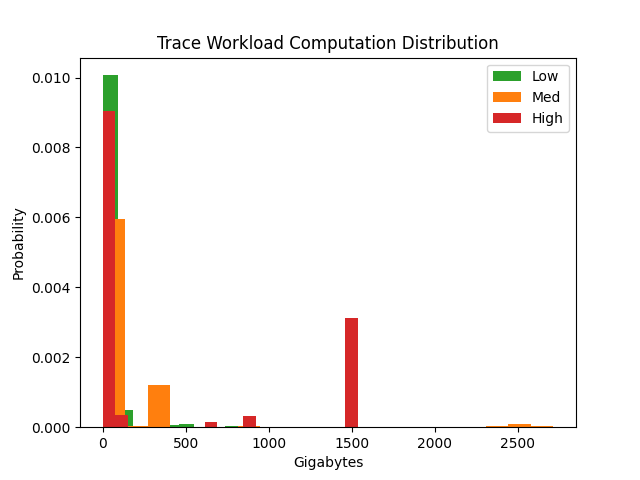}
\caption{Computation distribution from trace data.}
\label{fig:trace_computation}
\vspace{-14pt}
\end{figure}

\begin{figure}[t]
\includegraphics[width=8cm]{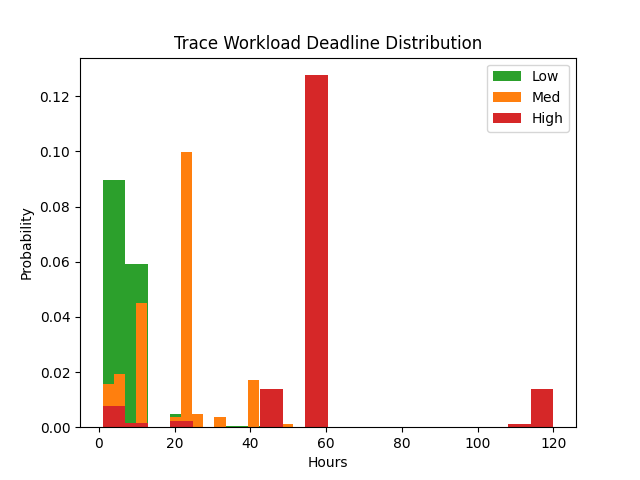}
\caption{Deadline distribution from trace data.}
\label{fig:trace_deadline}
\vspace{-14pt}
\end{figure}

The KnapsackGreedy algorithm was also tested using a trace based on real workload data from the University of Southampton. This trace has been used to evaluate previous work that follows the batch paradigm \cite{Rublein2022}. The full trace contains exact job arrivals and attributes from the past 4 years. A time window of 3 days in April 2021 was chosen as a representative sample of a steady workload (since traffic significantly decreased outside of the school semester). This time window was scaled into discrete timesteps such that each server ran an auction on all jobs every 10 minutes. The submitted jobs are divided into high, medium, and low priority categories based on user group (e.g., jobs submitted by faculty have higher priority than those submitted by undergraduates). Detailed histograms of trace job properties are shown in Figs.~\ref{fig:trace_storage}-\ref{fig:trace_deadline}.

Server sizes were set based on real nodes in the cluster. To generate some congestion, a small sample of nodes were used: two high-memory ($768$\:GB RAM) nodes and three regular ($192$\:GB RAM) nodes. These statistics were used for the servers' storage and computation, and the servers' download capacity$\times$slot duration $B_{d,i}$ was distributed normally with $\mu = 10$\:GB and $\sigma = 0.2$\:GB.

\begin{figure}[t]
\includegraphics[width=8cm]{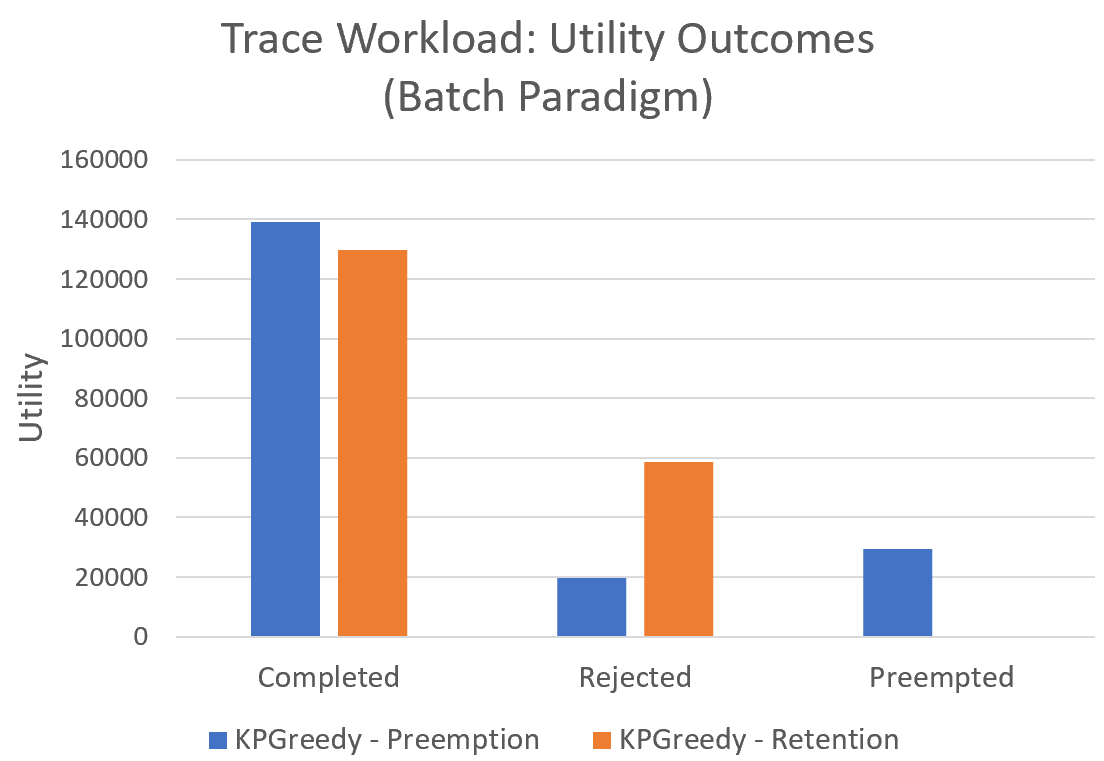}
\caption{Utility outcomes (completed, rejected, preempted) for Knapsack Greedy with and without preemption using the trace workload.}
\label{fig:traceUtility}
\end{figure}

The result of running our algorithms on the trace data is shown in \ref{fig:traceUtility}. For comparison, the Double Knapsack achieved $\approx$143,000 utility ($\approx$4000 more than KnapsackGreedy with preemption) \cite{Rublein2022}. Similar to the synthetic data results, the preemptive version of KnapsackGreedy achieves a slightly higher utility than its non-preemptive counterpart. In terms of runtime, Double Knapsack averages about $10$ minutes per trace auction, whereas KnapsackGreedy with and without preemption average about $3$ and $2$ minutes per auction, respectively.

\begin{figure}[t]
\includegraphics[width=8cm]{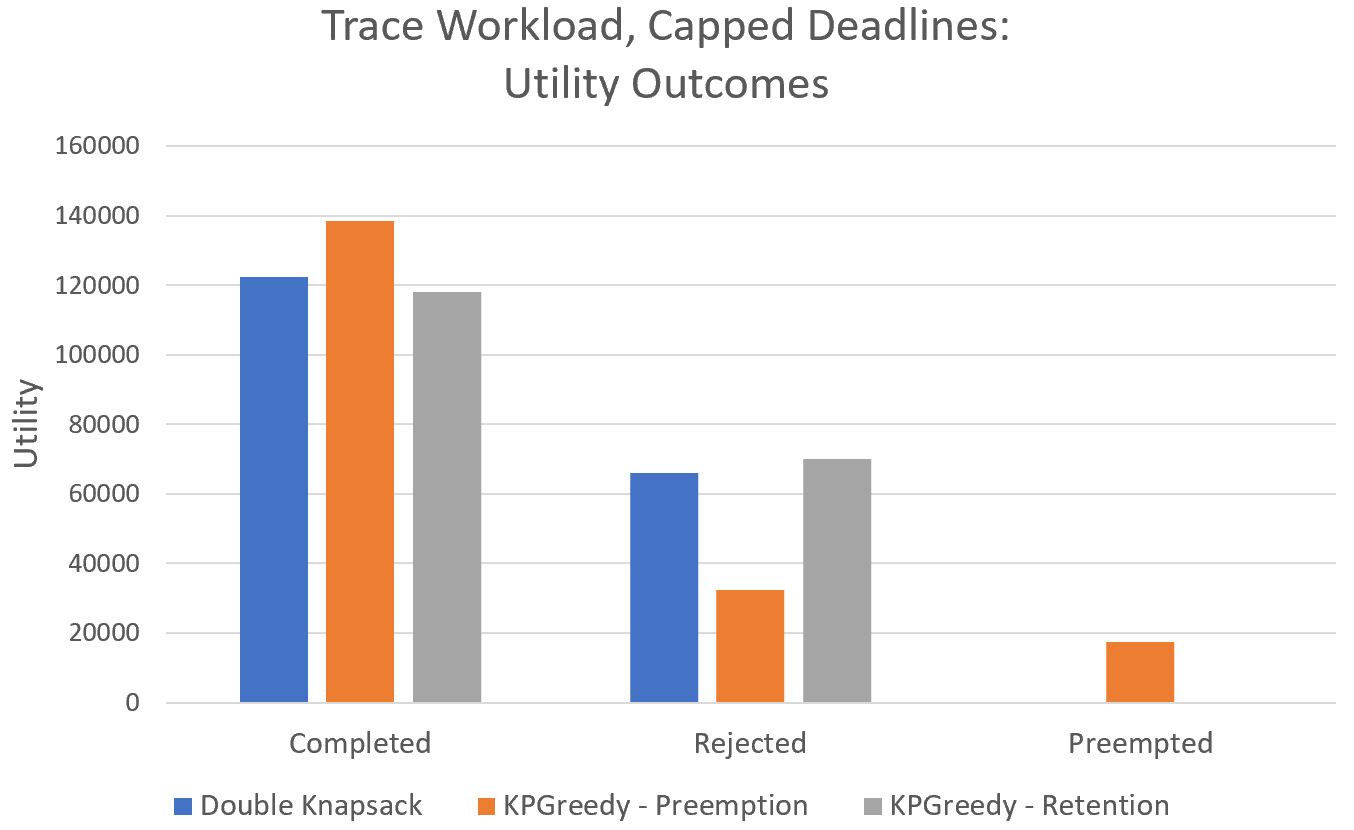}
\caption{Utility completed by each system when trace job deadlines are capped to 2 hours maximum.}
\label{fig:traceCappedDeadlines}
\end{figure}

We also observed the system when the deadlines of the trace jobs were limited to $2$ hours maximum. As shown in Fig.~\ref{fig:traceCappedDeadlines}, KnapsackGreedy outperforms Double Knapsack when the trace deadlines are limited in length. When auction times are taken into consideration, the double knapsack takes too long to run the bidding process, hindering its performance. KnapsackGreedy does well with short deadlines, but similar to other faster methods \cite{Rublein2022}, improves beyond Double Knapsack only with the help of preemption. Thus, \textit{KnapsackGreedy with preemption achieves the best balance between performance and auction time}.

\section{Conclusion}
\label{Conclusion}
In this paper, we present an optimal formulation for a preemption-enabled, edge cloud task allocation system in which tasks follow a pipeline paradigm. We also utilize a greedy heuristic as a scalable alternative that maintains the quality of preemption decisions. Both simulated workloads and real-world trace data were employed to thoroughly test this heuristic against other algorithms.
Our results show one algorithm with near-optimal performance and another that can achieve similar results \textcolor{black}{up to $2$-$5\times$ faster} for certain workloads.
Future work will include learning at the client side for more realistic user behavior.

\section*{Acknowledgment}
This research was sponsored by the U.S. Army Research Laboratory and the U.K. Ministry of Defence under Agreement Number W911NF-16-3-0001. The views and conclusions contained in this document are those of the authors and should not be interpreted as representing the official policies, either expressed or implied, of the U.S. Army Research Laboratory, the U.S. Government, the U.K. Ministry of Defence or the U.K. Government. The U.S. and U.K. Governments are authorized to reproduce and distribute reprints for Government purposes notwithstanding any copyright notation hereon.

The authors also acknowledge the support from University of Southampton's High Performance Computing Team regarding their help on providing the trace data.

\bibliographystyle{ieeetr}
\bibliography{references.bib}

\begin{IEEEbiography}
[{\includegraphics[width=1in,height=1.25in,clip,keepaspectratio]{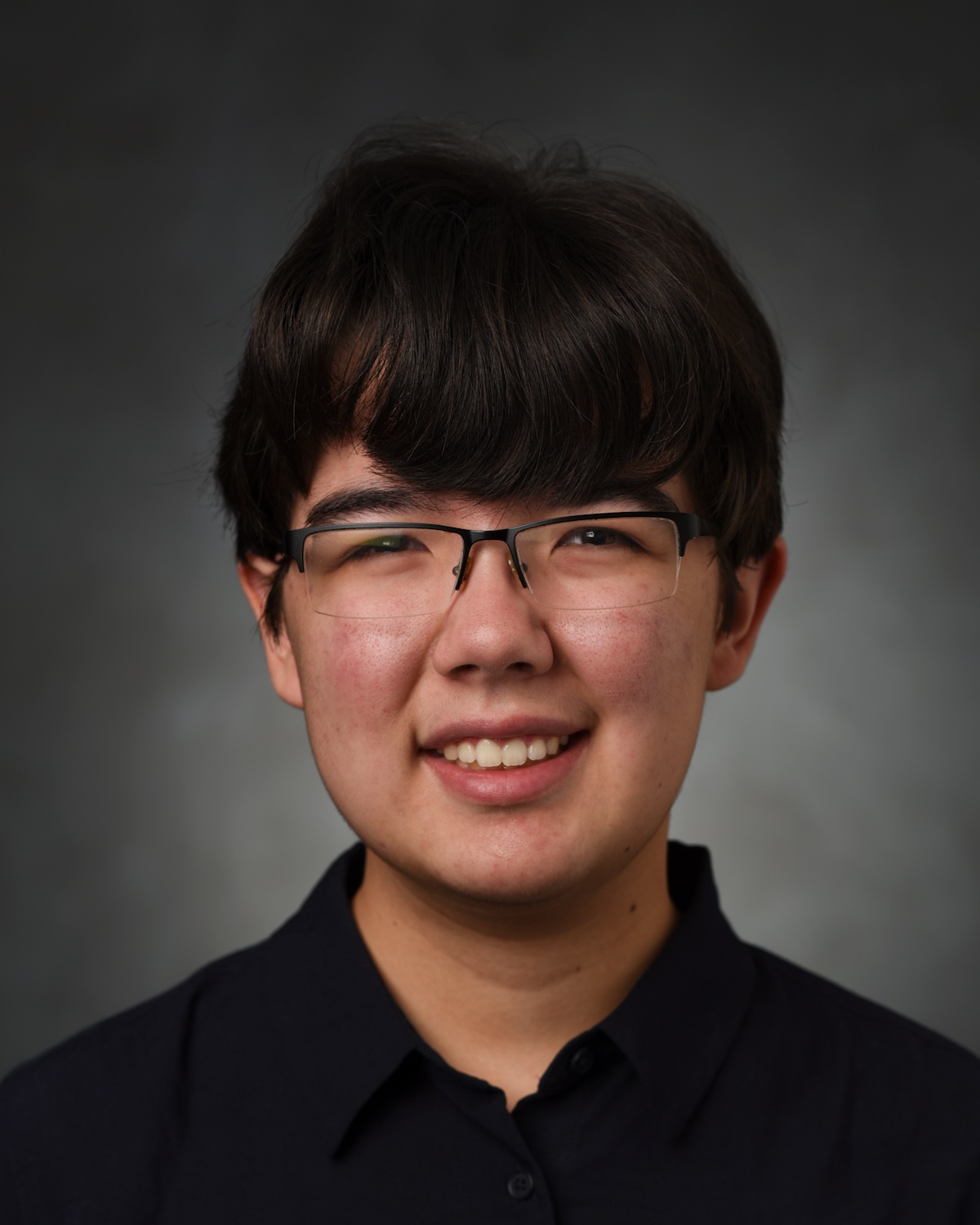}}]
{Caroline Rublein} 
is a PhD student in the Computer Science \&  Engineering  Department at Penn State University (USA), and is currently researching resource allocation in  networks. They received their B.S. in Applied Computer Science from Lock Haven University (USA) in 2019.
\end{IEEEbiography}

\begin{IEEEbiography}
[{\includegraphics[width=1in,height=1.25in,clip,keepaspectratio]{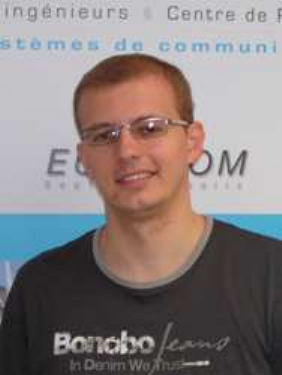}}]{Fidan Mehmeti} 
received his graduate degree in Electrical and Computer Engineering from the University of Prishtina, Kosovo, in 2009. He obtained his Ph.D. degree in 2015 at Institute Eurecom/Telecom ParisTech, France. After that, he was a Post-doctoral Scholar at the University of Waterloo, Canada, North Carolina State University and Penn State University, USA. 
He is now working as a Senior Researcher and Lecturer at the Technical University of Munich, Germany. His research interests lie within the broad area of wireless networks, with an emphasis on performance modeling, analysis, and optimization.  
\end{IEEEbiography}

\begin{IEEEbiography}
[{\includegraphics[width=1in,height=1.25in,clip,keepaspectratio]{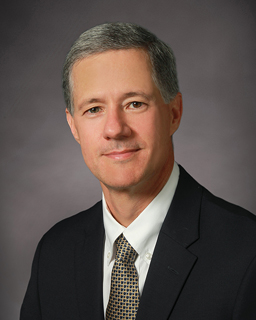}}]
{Mark Mahon} 
is a Teaching Professor in the School of Electrical Engineering and Computer Science at Penn State University. He received his B.S.E.E. from the University of Scranton and his M.S.E.E. and Ph.D. in Acoustics from the Pennsylvania State University. In 1991 he joined Penn State’s Applied Research Laboratory as a
research faculty member working primarily on various DoD research projects in the area of wireless communications. In 2015 he joined the School of EECS as a teaching faculty member.
Dr. Mahon’s primary area of research is cellular and mobile networks, signal processing, cybersecurity, and neural networks.  
\end{IEEEbiography}

\begin{IEEEbiography}
[{\includegraphics[width=1in,height=1.25in,clip,keepaspectratio]{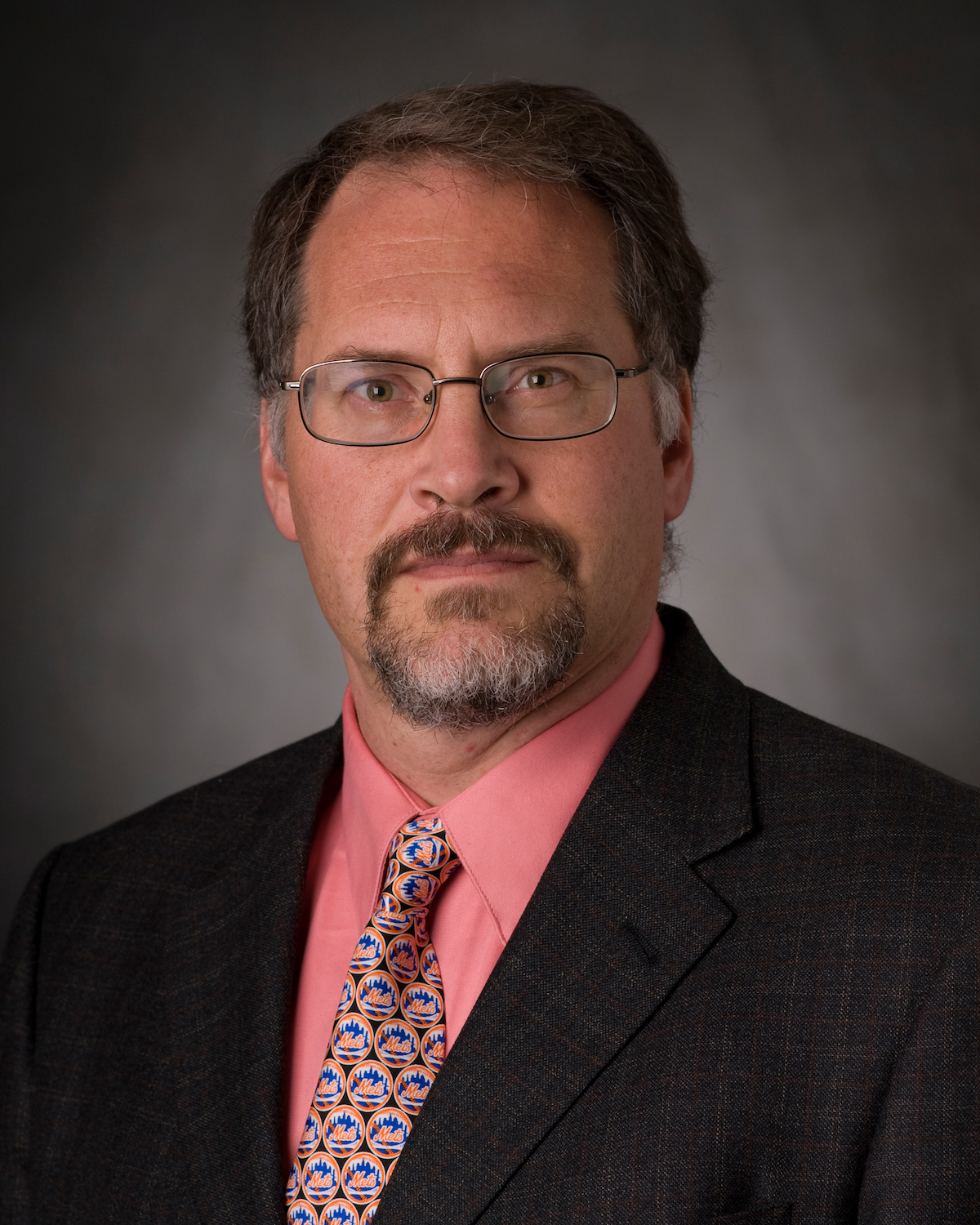}}]
{Thomas F. La Porta} 
is the Director of the School of Electrical Engineering and Computer Science at Penn State University. He is an Evan Pugh Professor and the  William  E.  Leonhard Chair Professor in the Computer Science and Engineering Department and the Electrical Engineering Department.  He received his B.S.E.E. and M.S.E.E. degrees from The Cooper Union, New York, NY, and his Ph.D. degree in Electrical Engineering from Columbia University, New York,NY. He joined Penn State in 2002. He was the founding Director of the Institute of Networking and Security Research at Penn State. Prior to joining Penn State, Dr. La Porta was with Bell Laboratories for 17 years. He was the Director of the Mobile Networking Research Department in Bell Laboratories, Lucent Technologies where he led various projects in wireless and mobile networking.  He is an IEEE Fellow, Bell Labs Fellow, and received the Bell Labs Distinguished Technical  Staff Award.  He also won two Thomas Alva Edison Patent Awards.  Dr. La Porta  was the founding Editor-in-Chief of the IEEE Transactions on Mobile Computing. He has published numerous papers and holds 39 patents.
\end{IEEEbiography}

\end{document}